\documentclass{ws-ijmpb}
\renewcommand{\theequation}{\arabic{section}.\arabic{equation}}

\def\be{\begin{equation}}
\def\ee{\end{equation}}
\def\bqa{\begin{eqnarray}}
\def\eqa{\end{eqnarray}}
\def\roughly#1{\mathrel{\raise.3ex
\hbox{$#1$\kern-.75em\lower1ex\hbox{$\sim$}}}}
\def\lsim{\roughly<}

\begin{document}

\markboth{Dieter Schildknecht}
{The Saturation of the Infrared Absorption by Carbon Dioxide in the
Atmosphere}

%
\catchline{}{}{}{}{}
%

\title{SATURATION OF THE INFRARED ABSORPTION\\
BY CARBON DIOXIDE IN THE ATMOSPHERE}

\author{DIETER SCHILDKNECHT}

\address{Fakult\"{a}t f\"{u}r Physik, Universit\"{a}t Bielefeld \\ 
D-33501 Bielefeld, Germany\\
schild@physik.uni-bielefeld.de }

\maketitle

\begin{history}
\received{Day Month Year}
\revised{Day Month Year}
\end{history}

\begin{abstract}
Based on new radiative transfer numerical evaluations, we reconsider an
argument presented by Schack in 1972 that says that saturation of
the absorption of infrared radiation by carbon dioxide in the atmosphere
sets in as soon as the relative concentration of carbon dioxide
exceeds a lower
limit of approximately 300 ppm. We provide a concise brief and explicit
representation of the greenhouse effect of the earth's atmosphere.
We find   an equilibrium climate sensitivity (temperature increase  
$\Delta T$  due to doubling of atmospheric $CO_2$  
concentration) of  $\Delta T  \simeq  0.5 ^0C$. We elaborate on the consistency of these results on $\Delta T$ with results observationally obtained by satellite-based measurements of short-time radiation-flux versus surface-temperature changes. 
\end{abstract}

\keywords{Greenhouse effect; Radiative transfer; Equilibrium climate.} 

\section{Introduction}

In view of the immense literature concerning the effect of an
(anthropogenic) increase in the relative atmospheric carbon dioxide
concentration on the
earth's climate, it is  appropriate to revive a simple
argument on this subject presented by Schack\cite{Schack_PB} in 1972.
The three-page
article published in German in the journal of the German Physical Society,
at that time called ``Physikalische Bl\"atter'', appeared under the title
``Der Einflu{\ss} des Kohlendioxid-Gehalts der Luft auf das Klima der Welt'',
i.e. ``The influence of the carbon dioxide content of the air on the climate
of the world''. The author, Schack, an expert on heat transfer
in combustion devices, in 1923 had discovered the role played
by the interaction of infrared electromagnetic radiation with carbon dioxide,
$CO_2$, and water vapor, $H_2O$, in such devices\cite{Schack}.

The intensity of (monochromatic) radiation transversing an
absorbing medium decreases
exponentially with distance according to what is known as Lambert-Beer's
law,
\be
I(z) = I (z=0) e^{-\kappa z}.
\label{1}
\ee
Independently of the numerical value of the absorption constant $\kappa$ in
(\ref{1}), for sufficiently large distance, $z$, the limit of ``complete
absorption'', or zero intensity of the outgoing radiation, $I (z \to \infty)
\to 0$, is reached. For finite values of the exponent $-\kappa z$ in (\ref{1}),
the numerical value of the $\kappa$ determines the distance,
$z_\epsilon (\kappa)$, for ``approximate saturation'',
$\exp (- \kappa z_\epsilon (\kappa))=\epsilon \ll 1$,
or equivalently, ``almost full absorption'',
\be
\frac{I(z=0) - I(z=z_\epsilon (\kappa))}{I(z=0)} =
1 - \frac{I(z=z_\epsilon(\kappa))}
{I(z=0)} = 1 - \epsilon.
\label{2}
\ee

In the case of $CO_2$ in air, the wide band absorption constant $\kappa$ for
the infrared
electromagnetic radiation depends on the concentration, or the partial
pressure, of the $CO_2$, and a natural question concerns the magnitude of
the $CO_2$ concentration that leads to approximate saturation within
the troposphere of the earth.

In his 1972 article\cite{Schack_PB}, Schack points out that for a
concentration of 0.03 \%
carbon dioxide in air, approximate saturation is reached within a distance
of approximately the magnitude of the height of the troposphere. The
absorption reaches values close to 100 \% for a realistic $CO_2$ content
of 0.03 \%, it is concluded\cite{Schack_PB} that any further increase
of (anthropogenic) $CO_2$ cannot lead to an appreciably stronger
absorption of radiation, and consequently cannot affect the earth's climate.

It will be useful to elaborate on the argument given by Schack in detail,
in order to explicitly display the simplicity and generality of the
underlying concepts that lead to a parameter-free prediction of the
absorption of infrared radiation by $CO_2$.

Adopting Planck's radiation law for a temperature at the surface of
the earth, chosen as $T = 293 K$ by Schack, and taking into account
the well-known absorption spectrum of the $CO_2$ molecule,
one finds that the radiation of
wave lengths $\lambda_{CO_2}$ in the interval
$13~ \mu m \le \lambda_{CO_2} \le 17.6~ \mu m$ is relevant for the absorption
by $CO_2$. The total absorption due to $CO_2$ in the atmosphere is
determined by the total mass of $CO_2$ that is transversed by a beam of
infrared radiation on its path from the surface at $z = 0$ to the upper
end of the atmosphere, or $z \to \infty$.

In the gravitational field of the earth, the pressure, $p$, of a gas
decreases with increasing altitude, $z$, according to $dp = - \rho
g~dz$, where $\rho$ denotes the density of the gas and $g$ the
acceleration due to gravity. For an ideal gas of temperature $T$, we have
$p = \rho RT/M$, or $\rho = pM/RT$, with $R$ being the gas constant, 
$T$ denoting the absolute temperature and $M$ the molecular weight of the
gas. The total mass per unit area transversed by a beam of infrared
radiation on its path through the atmosphere is determined by an
integration over the density $\rho (p,T)$ from the surface to the upper
end of the troposphere. The result of the integration may be represented
in terms of an effective altitude $z_0$ of a fictitious atmosphere of
homogeneous constant pressure $p_0$, constant temperature $T$ and constant
density $\rho$. The value of $z_0$ (obviously) depends on whether the
atmosphere is treated isothermally, or rather more realistically, is described
adiabatically.

Adopting
$T = {\rm const}$, the decrease of pressure with increasing altitude
becomes equal to
\be
p(z) = p_0 \exp \left(- \frac{Mg}{RT} z \right),
\label{3}
\ee
which is the well-known barometric formula. The density $\rho = \rho (z)$
decreases accordingly as
\be
\rho (z) = \frac{M}{RT} p_0 \exp \left(- \frac{z}{z_0}\right) \equiv \rho_0
\exp \left( - \frac{z}{z_0}\right),
\label{4}
\ee
where
\be
z_0 = \frac{RT}{Mg}
\label{5}
\ee
denotes the scale-height of the atmosphere, the height of a fictitious
atmosphere of homogeneous pressure $p_0$, temperature $T$ and density $\rho_0$,
and a mass per
unit area,
$M_0$, obtained from (\ref{4}),
\be
M_0 \equiv \int^{\infty}_{0} \rho (z) dz = \rho_0 z_0 = \frac{Mp_0}{RT}
z_0.
\label{6}
\ee
The determination of the absorption of the atmosphere is accordingly
reduced
to the evaluation of the absorption by a gas pipe of length $z_0$
from (\ref{5})
homogeneously filled with air, and with $CO_2$ of partial pressure
$p_0 \equiv p_{CO_2}$, at a total pressure of 1 atm at temperature
$T = 293~K$.

For the realistic case of an adiabatically treated atmosphere relation
(\ref{6}) remains valid with an appropriately chosen value of $z_0$
different from (\ref{5}). In our approach of Section 2 the value of
$z_0$ will be determined from the altitude dependence of the
absorption of infrared radiation for given lapse rate in the atmosphere.

The absorption due to (\ref{1}), upon introducing the
dependence on the $CO_2$ concentration via the partial pressure
$p_{CO_2}$, becomes
\be
A (p_{CO_2}, z_0) = 1 - \exp \left( -a p_{CO_2}z_0 \right),
\label{7}
\ee
where the coefficient $a$ must be determined from the known
absorption spectrum of $CO_2$. Since the absorption cross section of
$CO_2$ depends on the wave length of the radiation being absorbed,
the absorption constant $a$ in (\ref{7}), obtained by summation over a spectrum
of a large number of spectral lines,  actually develops a
dependence on the partial pressure $p_{CO_2}$.

\begin{table}[h]   
\tbl{The results given in the first four horizontal lines are taken
from ref. $^1$,
the results in the last line are computed
using (\ref{7}).}
{\begin{tabular}{@{}lcccc@{}} \Hline 
\\[-1.8ex] 
$CO_2$ [\%] & 0.03 \% & 0.06 \% & $ \Delta_{2 \times CO_2}~[\%]$
& $ \Delta_{2 \times CO_2}[Wm^{-2}]$ \\
$p_{CO_2} z_0 [atm \times m]$ & 2.1 & 4.2 & & \\
A [\%] & 98.5 \% & 99.3 \% & 0.8 \% & 0.70\\
1 - A [\%] & 1.5 \% &0.7 \% & &\\
$a [(atm \times m)^{-1}]$ & 2.0 & 1.73 & &\\ 
\Hline \\[-1.8ex] 
\end{tabular}}
\end{table}

In his article\cite{Schack_PB}, addressing the question on the effect
of an increase of the $CO_2$
concentration on the climate on earth,
Schack compares the absorption due to $CO_2$, 
by a $CO_2$-air pipe
as mentioned, for a partial pressure
corresponding to a realistic value of 300 ppm = 0.03 \% with the absorption
due to an enhanced value taken as
\break 600 ppm = 0.06 \% \footnote{ A realistic
value for the increase of the $CO_2$ concentration per year is 2 ppm/year, e.g.\cite{Gervais}, 
implying 150 years for the doubling of a $CO_2$ concentration
of 300 ppm.}, respectively
corresponding to $p_{CO_2} = 3 \times 10^{-4}$ atm and $p_{CO_2} =
6 \times 10^{-4}$ atm  at T = 293 K and
a total pressure of 1 atm of the atmosphere. Adopting
$z_0 = 7000~m$ for the equivalent height of the troposphere, for the absorption,
A [\%], and the increase of absorption, $\Delta_{2 \times CO_2}$,
Schack obtains the
results that are given in Table 1.

Saturation of infrared radiation by the realistic $CO_2$ content of the
atmosphere of\break 0.03 \% being reached within approximately 1 \%
according to Table 1, any further
increase of the $CO_2$ content
does not substantially increase the absorption of
radiation, and accordingly does not affect\cite{Schack_PB} the
climate on earth.

The results on $\Delta_{2 \times CO_2}~[\%]$ in Table 1 refer to
the radiation in the $CO_2$ band of 13 $\mu m \lsim \lambda_{CO_2}
\lsim 17.6 \mu m~ (568 cm^{-1} \lsim 1/\lambda_{CO_2} \lsim 769
{\rm cm}^{-1}).$ With the assumption of a black-body radiation at
$T = 293 K$, the input radiation in the $CO_2$ band amounts to
$87 W/m^2$. The additional absorption due to $CO_2$ doubling is
given by $\Delta_{2 \times CO_2} \times 87 W/m^2 = 0.8 \times
10^{-2} \times 87 W/m^2 = 0.70 W/m^2$, as listed in Table 1.
This increase of the absorption of infrared radiation in the
atmosphere is indeed negligible in comparison with the equilibrium
value of the total outgoing long-wave infrared radiation (LWIR) at
the top of the atmosphere (TOA) given by $S_0 = 239 W/m^2$.
\footnote{The known empirical
value of the sun radiation of
1365 $W/m^2$ with an albedo of 0.3 yields 956 $W/m^2$ absorption. Dividing
by 4 to take into account the surface of the earth yields the average value
of the radiation from the sun to be balanced by the outgoing average LWIR
of $S_0 = 239~ W/m^2$. \cite{Trenberth} A 1 \% decrease of the albedo to become $\epsilon = 0.297$ yields $S_0 = 240 W/m^2$ to be used subsequently.}

\section{Absorption of Infrared Radiation: $CO_2$-Air Pipe versus Atmosphere}

Using presently available detailed data on the spectral absorption lines
of $CO_2$, compare\cite{HITRAN}, the absorption coefficient\footnote{Compare
e.g. ref.\cite{Wei} for a recent evaluation of absorption coefficients}
a and the
resulting absorption by a $CO_2$-air pipe, defined before, can be reliably
predicted. In this Section, we compare the results on absorption by a
$CO_2$-air pipe at fixed temperature and pressure, as a function of the
length of the $CO_2$-air pipe, with the absorption by an adiabatically
treated atmosphere. The results to be presented are due to a recent
evaluation by Clark\cite{Clark} for the $^{12}C~^{16}0_2$ molecule
only, with linestrengths $> 10^{-23}$. We refer to Appendix A for further
details.

\begin{table}[h]   
\tbl{$CO_2$-air pipe absorption A and increase $\Delta_{2 \times CO_2}$
for selected path lengths and $CO_2$ concentrations for the spectral range
568 to 769 cm$^{-1}$ (input black-body radiation 87 $Wm^{-2}$) and
$T = 293 K,~p = 1 atm$. Compare Appendix A Table A2.}
{\begin{tabular}{@{}lcccc@{}} \Hline 
& A	& A & &  \\
& 300 ppm & 600 ppm & $\Delta_{2 \times CO_2}$ [ \%] & $\Delta_{2 \times CO_2}
[Wm^{-2}]$ \\
\Hline
1 km & 0.63 & 0.71 & 8 \% & 7.00 \\
2 km & 0.71 & 0.78 & 7 \% & 6.09 \\
3 km & 0.75 & 0.82 & 7 \% & 6.09 \\
5 km & 0.81 & 0.87 & 6 \% & 5.22 \\
7 km & 0.84 & 0.89 & 5 \% & 4.35 \\
\Hline
\end{tabular}}
\label{tab2a}
\end{table}

\begin{table}[h]   
\tbl{Same as Table 2, but for 600 cm$^{-1}$ to 750 cm$^{-1}$ spectral
range (input black-body radiation 65 $Wm^{-2}$). Compare Appendix A Table A3.}
{\begin{tabular}{@{}lcccc@{}} \Hline 
\\[-1.8ex] 
& A	& A & &  \\
& 300 ppm & 600 ppm & $\Delta_{2 \times CO_2}$ [ \%] & $\Delta_{2 \times CO_2}
[Wm^{-2}]$ \\
\Hline
1 km & 0.79 & 0.87 & 8 \% & 5.20 \\
2 km & 0.87 & 0.93 & 6 \% & 3.90 \\
3 km & 0.91 & 0.96 & 5 \% & 3.25 \\
5 km & 0.95 & 0.98 & 3 \% & 1.95 \\
7 km & 0.97 & 0.99 & 2 \% & 1.30 \\
\Hline
\end{tabular}}
\label{tab2b}
\end{table}

The results\cite{Clark} for the absorption of a $CO_2$-air pipe for the
spectral range of $13 \mu m \lsim \lambda_{CO_2} \lsim 17.6 \mu m$
(568 cm$^{-1}$ to 769 cm$^{-1}$), as a function of selected values of
the path length (pipe length) and the relevant $CO_2$ concentrations
of 0.03 \% and 0.06 \% are shown in Table 2. The pipe is assumed to have
constant temperature, $T = 293~K$, and a total pressure of $p = 1 atm$. The
input black-body infrared radiation at $T = 293~K$ is accordingly
given by $87 Wm^{-2}$. The results in Table 2 explicitly demonstrate
the expected increase of the absorption A with increasing length of
the pipe (path length), as well as the decrease of the relative
absorption for $CO_2$ doubling, $\Delta_{2 \times CO_2}~[\%]$. The
result of $\Delta_{2 \times CO_2} = 5 \% $ at the pipe length of 7 km is
considerably larger than
the 1972 result of $\Delta_{2 \times CO_2} = 0.8 \% $ given in Table 1.

\begin{figure}[t]
\centerline{\psfig{file=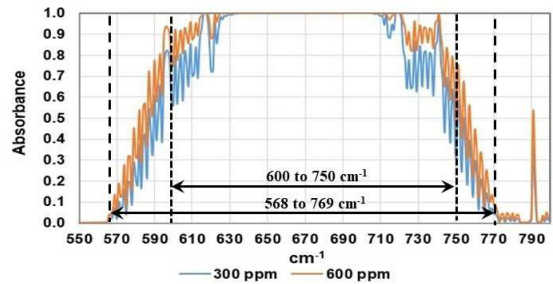,width=3in}}
\caption{The absorbance of 0.03 \% and 0.06 \% concentrations of $CO_2$
in dry air at 293 K for a 3 km $CO_2$-air pipe. The spectral ranges of
568 to 769 cm$^{-1}$ 
and 600 to 750 cm$^{-1}$ 
used in the
absorption band calculations are indicated. The calculation $^7$ was
performed at a spectral resolution of 0.01 cm$^{-1}$, but the
data are plotted at 1 cm$^{-1}$.}
\end{figure}

In Table 3, for illustration, we show the effect of restricting
the spectral range of the input black-body radiation to the spectral
range of $600~cm^{-1}$ to $750~cm^{-1}$ (total black-body radiation
$65~Wm^{-2}$). Comparing the results in Table 3 with the ones
in Table 2, we notice the more rapid approach to a (larger) asymptotic
absorption limit with increasing path length of the infrared radiation.
This is a consequence of the change of absorbance as a function of the
spectral wave length taken into account.

In Fig. 1, we
show the calculated spectrum of the absorbance for the case
of a path length of 3 km and a $CO_2$ concentration of $0.03 \%$ and
$0.06 \%$. The spectral ranges corresponding to Tables 2 and 3 are
indicated in Fig. 1.

Turning from fixed temperature $T$ in the $CO_2$-air pipe to decreasing
$T$ and pressure $p$ in the atmosphere, we proceed in two steps. In a
first step, modelling a dry atmosphere,
we consider the effect of $CO_2$ (at concentrations of
$0.03 \%$ and $0.06 \%$), and in a second step, we add water vapor of
fixed relative humidity (RH) chosen as $RH = 85 \%$.

\begin{table}[h]   
\tbl{Altitude dependence of the infrared absorption A by the atmosphere
for selected altitudes and $CO_2$ concentrations for the 568 to 769 cm$^{-1}$
spectral range (input black-body radiation 87 $Wm^{-2}$). Surface temperature
and surface pressure are T = 293 K and p = 1 atm, and the lapse rate adopted
for the atmosphere is given by $-6.5~^0C km^{-1}$. Compare Appendix A Table A4.}
{\begin{tabular}{@{}lcccc@{}} \Hline 
\\[-1.8ex] 
& A & A & &  \\
& 300 ppm & 600 ppm & $\Delta_{2 \times CO_2}$ [ \%] & $\Delta_{2 \times CO_2}
[Wm^{-2}]$ \\
\Hline
Atm 5 km & 0.74  & 0.81 & 7 \%  & 6.09  \\
Atm 9 km & 0.76 & 0.83 & 7 \% & 6.09 \\
Atm 11 km & 0.76 & 0.83 & 7 \% & 6.09 \\
\Hline 
\end{tabular}}
\label{tab3a}
\end{table}

\begin{table}[h]   
\tbl{Same as Table 4, but for a 600 to 750 $cm^{-1}$ spectral range
(input black body emission 65 $Wm^{-2}$).Compare Appendix A Table A4.}
{\begin{tabular}{@{}lcccc@{}} \Hline 
\\[-1.8ex] 
& A & A & & \\
& 300 ppm & 600 ppm & $\Delta_{2 \times CO_2}$ [ \%] & $\Delta_{2 \times CO_2}
[Wm^{-2}]$ \\
\Hline
Atm 5 km & 0.90  & 0.95  & 5 \%  & 3.25  \\
Atm 9 km & 0.91 & 0.96 & 5 \% & 3.25 \\
Atm 11 km & 0.91 & 0.96 & 5 \% & 3.25 \\
\Hline
\end{tabular}}
\label{tab3b}
\end{table}

In Table 4, we show the results for the absorption A of the atmosphere
at several selected altitudes for a $CO_2$ content of $0.03 \%$ and $0.06 \%$.
The spectral range of the LWIR of $568~cm^{-1}$ to $769~cm^{-1}$ is
identical to the spectral range employed for the results in Table 2. The
evaluation of the absorption was carried out for a surface temperature
of $T = 293~K$ and a surface pressure $p = 1 atm$, and a realistic
moist adiabatic lapse rate increasing with altitude, and an average
value of $-6.5~^0C~km^{-1}$ was employed (compare Appendix A, Fig. A6). 
The absorption spectrum at 0.01 $cm^{-1}$ resolution was calculated using a spatial resolution of 100 m. The absorption spectrum for each 100 m level was calculated using the pressure and temperature for that level derived from the lapse rate.
The results
in Table 4 show a rapid convergence to an approximately constant,
quasi-asymptotic behavior of the absorption $A$ that is reached above an
altitude of approximately 5 km. By comparison of the results in Table 4
with the ones in Table 2, we conclude that a $CO_2$-air pipe for a chosen
length of 2 to 3 km represents the absorption of the
atmosphere that reaches a quasi-asymptotic limit at altitudes of about 5 km.
The values of the absorption
and the increase of the absorption under $CO_2$ doubling of $\Delta_{2 \times
CO_2} = 7 \%$ corresponding to $6 Wm^{-2}$ in Tables 2 and 4 are in
agreement with each other. Similar conclusions are drawn from the comparison
of the results for the restricted spectral range in Tables 3 and 5 with
each other.

\begin{figure}[t]
\centerline{\psfig{file=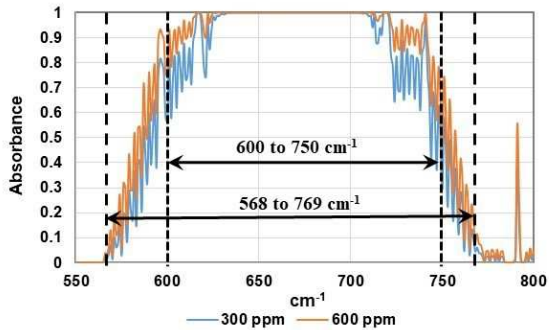,width=3in}}
\caption{The absorbance of 0.03 \% and 0.06 \% concentrations of $CO_2$
in a dry atmosphere for a vertical path length from the surface to 7 km
altitude. The surface temperature is 293 K. The lapse rate is - 6.5 K
${\rm km}^{-1}$ and the spatial resolution is 100 m. The calculation
$^7$
was performed at a spectral resolution of 0.01 ${\rm cm}^{-1}$, but the
data are plotted at 1 ${\rm cm}^{-1}$ resolution.}
\end{figure}

In Fig. 2, we show the absorption of the surface emission at
293 K of 300 and 600 ppm of $CO_2$ for the dry atmosphere
looking downwards from a height of 7 km. The total spectral
range is from 550 to 800 $cm^{-1}$ and the spectral intervals are indicated. These are the same as for Fig. 1. The spectral resolution of the calculation was 0.01 $cm^{-1}$ with a spatial resolution of 100 m. The spectral resolution of the plot however, is 1 $cm^{-1}$. This spectrum includes the effects of the decrease in temperature and pressure with altitude as shown in Appendix A, Fig. A6. In this case most of the absorption occurs near the surface and the value of the total absorption tends towards an asymptotic value as the altitude increases. This is shown in Appendix A, Fig. A4.

The results explicitly demonstrate that the $CO_2$-air pipe of path
length of approximately 3 km represents the absorption in the atmosphere
of a quasi-asymptotic altitude reached at 5 km.

\begin{table}[h]   
\tbl{Same as Table 4, but including the absorption by water vapor
of relative humidity (RH) of 85 \%. Compare Appendix A, Table A5.}
{\begin{tabular}{@{}lcccc@{}} \Hline 
\\[-1.8ex] 
& A & A & &  \\
& 300 ppm & 600 ppm & $\Delta_{2 \times CO_2}$ [ \%] & $\Delta_{2 \times CO_2}
[Wm^{-2}]$ \\
\Hline
Atm 5 km &  &   &   &  \\
85 RH &\raisebox{1.5ex}[-1.5ex]{0.89} & \raisebox{1.5ex}[-1.5ex]{0.92}
& \raisebox{1.5ex}[-1.5ex]{3 \%} & \raisebox{1.5ex}[-1.5ex]{2.6} \\
Atm 9 km &  &  &  &   \\
85 RH & \raisebox{1.5ex}[-1.5ex]{0.90} & \raisebox{1.5ex}[-1.5ex]{0.93}
& \raisebox{1.5ex}[-1.5ex]{3 \%} & \raisebox{1.5ex}[-1.5ex]{2.6} \\
Atm 11 km &  &  &   &  \\
85 RH & \raisebox{1.5ex}[-1.5ex]{0.90} & \raisebox{1.5ex}[-1.5ex]{0.93}
& \raisebox{1.5ex}[-1.5ex]{3 \%} & \raisebox{1.5ex}[-1.5ex]{2.6} \\ 
\Hline
\end{tabular}}
\label{tab4a}
\end{table}

\begin{table}[h]   
\tbl{Same as Table 5, but including the absorption by water vapor
of relative humidity (RH) of 85 \%. Compare Appendix A, Table A5.}
{\begin{tabular}{@{}lcccc@{}} \Hline 
\\[-1.8ex] 
& A & A & &  \\
& 300 ppm & 600 ppm & $\Delta_{2 \times CO_2}$ [ \%] & $\Delta_{2 \times CO_2}
[Wm^{-2}]$ \\
\Hline
Atm 5 km &  &  &  &  \\
85 RH & \raisebox{1.5ex}[-1.5ex]{0.94} & \raisebox{1.5ex}[-1.5ex]{0.97}
& \raisebox{1.5ex}[-1.5ex]{3 \%} & \raisebox{1.5ex}[-1.5ex]{2.0} \\
Atm 9 km &  & &  &  \\
85 RH & \raisebox{1.5ex}[-1.5ex]{0.95} & \raisebox{1.5ex}[-1.5ex]{0.98}
& \raisebox{1.5ex}[-1.5ex]{3 \%} & \raisebox{1.5ex}[-1.5ex]{2.0} \\
Atm 11 km &  &  &   &  \\
85 RH & \raisebox{1.5ex}[-1.5ex]{0.95} & \raisebox{1.5ex}[-1.5ex]{0.98}
& \raisebox{1.5ex}[-1.5ex]{3 \%} & \raisebox{1.5ex}[-1.5ex]{2.0} \\
\Hline
\end{tabular}}
\label{tab4b}
\end{table}

\begin{figure}[h] 
\centerline{\psfig{file=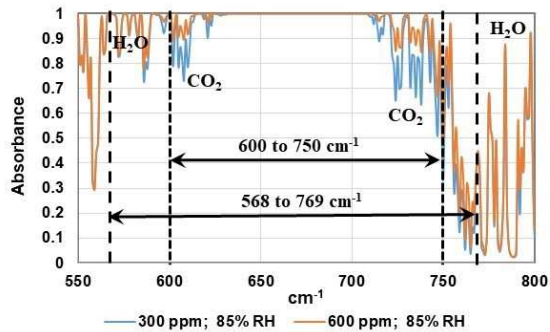,width=3in}}
\vspace*{8pt}
\caption{The absorbance of 0.03 \% and 0.06 \% concentrations of $CO_2$ in
a moist atmosphere for a vertical path length from the surface to 7 km
altitude. The surface temperature is 293 K. The relative humidity at the
surface is 85 \%. The lapse rate is - 6.5 K ${\rm km}^{-1}$ up to the
saturation level the spatial resolution is 100 m. The calculation $^7$ was
performed at a spectral resolution of 0.01 ${\rm cm}^{-1}$, but the
data are plotted at 1 ${\rm cm}^{-1}$ resolution.}
\end{figure}

The important role of water vapor -- semi-quantitatively discussed in
ref.\cite{Schack_PB} -- is illustrated by comparing the results in
Tables 4 and 5 with the results presented in Tables 6 and 7. The
results in Tables 6 and 7, in addition to $CO_2$, take into account
water vapor with relative humidity (RH) of RH = $85 \%$. The presence
of water vapor drastically reduces\footnote{Note that (obviously) the
presence of water vapor increases the absorption of radiation. The presence
of water vapor, however, decreases the effect of an increase of the
$CO_2$ concentration. Compare ref.\cite{Schack_PB} for a semi-quantitative
discussion.} the
effect of doubling the $CO_2$ concentration from $6.09~Wm^{-2}$ to
about $2.6~W~m^{-2}$ for the infrared spectral range of $568$ to
$769~cm^{-1}$. A similar effect is seen by comparing the results in
Tables 5 and 7 with each other.

For comparison with Fig. 2, in Fig. 3,
we show the absorption spectrum in the atmosphere for the presence of
water vapor in addition to carbon dioxide.

In summarizing this Section we conclude that a $CO_2$-air pipe of path
length of 2 to 3 km represents the effect of the $CO_2$ increase in
the atmosphere which has reached its quasi-asymptotic saturation limit
at an altitude of approximately 5 km.
Inclusion of water vapor drastically reduces the effect
on the absorption of a rise of
the $CO_2$ concentration from $0.03 \%$ to $0.06 \%$ from about $6 W~m^{-2}$
to $3 Wm^{-2}$. The value of $3 Wm^{-2}$ additional absorption
corresponding to $3/418 \simeq 0.7 \%$ of the total outgoing
radiation at the surface of the earth, we conclude that the doubling
of the $CO_2$ concentration for the relevant time scale of about a
century is a fairly negligible one. Even though the increase of
absorption of $\Delta_{2 \times CO_2} \simeq 3 Wm^{-2}$ is
considerably larger than the 1972 value of $\Delta_{2 \times CO_2}
\simeq 1 Wm^{-2}$, the main conclusion from ref.\cite{Schack_PB}
remains valid.

\section{$CO_2$ Doubling and Surface Temperature}
\renewcommand{\theequation}{\arabic{section}.\arabic{equation}}
\setcounter{equation}{0}

In Section 2, we concluded that the absorption of infrared radiation
from the $CO_2$ band in the atmosphere becomes (approximately) independent
of the altitude for altitudes larger than about 5 km; the absorption reaches
a quasi-asymptotic saturation limit at an altitude of about 5 km. The
magnitude of the quasi-asymptotic absorption agrees with the absorption
of a $CO_2$-air pipe at constant temperature, $T = 293 K$, and constant
pressure, $p = 1 atm$, and length of about 2 to 3 km.

The evaluation\cite{Clark}, compare Appendix A for details,
of the absorption by the atmosphere is based on the known spectroscopic
properties of the absorbing gases, $H_2O$, $CO_2$, and the empirically
known average lapse rate of the atmosphere of about $-6.5 K km^{-1}$.
The absorption reaches an approximately constant value at an
altitude of 6 km. At and beyond that height, a constant value of infrared radiation is freely emitted to space.
At this height, the atmosphere provides a ``free window'' (compare
Appendix B) for the emittance of infrared radiation at
temperature $T_{TOA} = (293 - 6 \times 6.5) K = 254 K \simeq 255 K$.
The outgoing intensity is given by $S_{space} = S(T) \vert_{T = 255 K} =
\sigma T^4 \vert_{T = 255} = 240~Wm^{-2}$.

The approximate stability of the climate on earth, a necessary condition
for life on earth, requires equilibrium of the outgoing infrared radiation,
predicted by $S_{space} = S(T) \vert_{T = 255 K}$, with the effective
radiation provided by the sun. The effective radiation by the sun amounts
\footnote{Compare footnote 3} to $S_0 = 240 Wm^{-2}$. The agreement of
the above parameter-free prediction of the radiative transfer calculation,
with the effective radiation from the sun,
\be
S_{space} = S(T) \vert_{T = 255 K} = S_0 ,
\label{3.0}
\ee
provides a non-trivial
support of the radiative-transfer calculation based on the empirical
lapse rate of $6.5 Km^{-2}$. Indeed, the radiative-transfer analysis
explains the approximate stability of the climate on earth.

For a concise representation of the radiative ``greenhouse effect''
in terms of the radiation from the sun, $S_0$, and the earth's LWIR within
as well as outside the ``atmospheric window'', as quantified by $S_W$,
and the fractional transmittance, $f^\prime \ll 1$, respectively,
compare Appendix B.

We turn to the effect of $CO_2$ doubling on the surface temperature. We
assume the existence of radiative equilibrium of the emitted black-body LWIR,
given by $S(T)$, for any given $CO_2$ content (within the presently
discussed limits). Absorption under doubling of the $CO_2$ concentration,
$\Delta_{2 \times CO_2}$, for compensation requires increase of radiation
$S (T) \to S (T) + \Delta S$, i.e.
\be
S(T) + \Delta S = S(T + \Delta T) = S(T) + \frac{dS}{dT}\bigg|_T \Delta T,
\label{3.1}
\ee
where $\Delta S \equiv \Delta_{2 \times CO_2}$ and $T$ denotes the surface
temperature, chosen as $T = 293 K$ in our treatment. Note that the
detailed mechanism occurring in the atmosphere leading to the required
increase of $\Delta S = \Delta_{2 \times CO_2}$ need not be specified. 
According to (\ref{3.1}), the temperature has to rise by
\be
\Delta T = \frac{\Delta_{2 \times CO_2}}{\left( \frac{dS}{dT}\right)_T}
= \frac{T \Delta_{2 \times CO_2}}{4 S(T)},
\label{3.2}
\ee
where $S(T) \sim T^4$ for a Planck radiator is used in the last step.
Relation (\ref{3.2}) specifies the rise of the surface temperature connected
with restoration of equilibrium at the end of a $CO_2$-air pipe, or
equivalently, at the altitude of the atmosphere where the quasi-asymptotic
limit of the absorption has been reached. Inserting $T = 293 K$, and
$S(T=293K) = 418 Wm^{-2}$, as well as
$\Delta_{2 \times CO_2} = 6.09 W~m^{-2}$ from
Tables 2 and 4 into (\ref{3.2}), we obtain an increase of surface
temperature of $\Delta T = 1.07~K$, compare Table 8. Insertion of
$\Delta_{2 \times CO_2} = 2.6 W~m^{-2}$ from Table 6 into (\ref{3.2})
shows the strong effect of water vapor of $RH = 85 \%$. The strong
absorption due to $H_2O$ diminishes the surface-temperature change to
$\Delta T = 0.46 K$ for the assumed increase of the
$CO_2$ concentration from $0.03 \%$ to $0.06 \%$.

\begin{table}[h]   
\tbl{The increase of surface temperature compensating infrared
absorption by $CO_2$ and by $CO_2 + H_20$ with $RH = 85 \%$}
{\begin{tabular}{@{}lcc@{}} \Hline 
\\[-1.8ex] 
& $\Delta_{2 \times CO_2}~[Wm^{-2}]$ & $\Delta T~[K]$ \\
\Hline
3 km $CO_2$-air pipe &  &  \\
9 km Atmosphere & \raisebox{1.5ex}[-1.5ex]{6.09}
& \raisebox{1.5ex}[-1.5ex]{1.07} \\
9 km Atmosphere & & \\
$RH = 85 \%$ & \raisebox{1.5ex}[-1.5ex]{2.6}
& \raisebox{1.5ex}[-1.5ex]{0.46} \\
\Hline
\end{tabular}}
\label{tab5}
\end{table}

\section{Radiative Absorbance under Change of Atmospheric Conditions.}
\renewcommand{\theequation}{\arabic{section}.\arabic{equation}}
\setcounter{equation}{0}

A very elaborate radiative transfer calculation, recently presented by Harde\cite{Harde}, 
within a two-layer climate model,
explores\footnote{Compare refs.\cite{Clark2,Chilinger} for alternative
dynamical climate models.}
the infrared absorbance by $CO_2$ under a variety of atmospheric conditions,
such as surface temperature, humidity, clouds etc. Results\cite{Harde}
on the radiative absorbance for $CO_2$ contents of 380 ppm and 760 ppm, under
variation of the atmospheric conditions, are shown in Table 9.

\begin{table}[h]   
\tbl{The first three columns quote the
results obtained by Harde $^{10}$  for the absorbance of
infrared radiation for $CO_2$ concentrations of 0.038 \% and 0.076 \% ,
and the increase $\Delta_{2 \times CO_2} [\%]$, under
different atmospheric conditions,
such as surface temperature, humidity, clouds etc. In the remaining
four columns, we present
the associated increase, $\Delta_{2 \times CO_2}~[Wm^{-2}]$,
and the surface temperature increase, $\Delta T$, and the averages
$\overline{\Delta}_{2 \times CO_2}$ and $\overline{\Delta T}$.}
{\begin{tabular}{@{}ccccccc@{}} \Hline 
\\[-1.8ex] 
&&&&&& \\
0.038 \% & 0.076 \% & $\Delta_{2 \times CO_2}$ [\%] &
$\Delta_{2 \times CO_2}$ & $\Delta T$
& $\overline{\Delta}_{2 \times CO_2}$ 
& $\overline{\Delta T}$ \\
& & & $[Wm^{-2}]$ & $[K]$ & $[Wm^{-2}]$ & $[K]$ \\
&&&&&& \\
\Hline 
82.57 \% & 83,84 \% & 1.27 \% & 5.03 & 0.92 & & \\
82.57 \% & 83.31 \% & 0.74 \% & 2.93 & 0.53 & 3.45 & 0.63\\
89.77 \% & 90.48 \% & 0.71 \% & 2.81 & 0.51 & &  \\
81.55 \% & 82.05 \% & 0.50 \% & 1.98 & 0.36 & & \\
84.08 \% & 85.22 \% & 1.14 \% & 4.51 & 0.82 & & \\
\Hline
\end{tabular}}
\label{tab6}
\end{table}

We note that the results
for the absorbance in Table 9 give the absorption due to $CO_2$ doubling
as a fraction of the total LWIR of $S(T) = 396~Wm^{-2}$ at $T = 289~K$ employed
by Harde, compared to Tables 2, 3 through 6, 7 that
give the results for the increase of absorbance from $CO_2$ doubling
as a fraction of the radiation in the $CO_2$ band. For a detailed
discussion, we refer to ref.\cite{Harde}.

In the context of the present discussion, the essential and important
point of the results in Table 9 is the approximate independence of the
absorption due to $CO_2$ doubling, $\Delta_{2 \times CO_2} \cong 1 \%$,
from the
various atmospheric conditions evaluated. Compare the results for
$\Delta_{2 \times CO_2}$ in column 3 of Table 9. The spread of the
results for $\Delta T$ in column 5 of Table 9 between $\Delta T =
0.36 K$ and $\Delta T = 0.92 K$, and the average value of $\overline{\Delta T}
= 0.63 K$, are consistent with the result from our straight-forward
evaluation of $\Delta T = 0.46 K$ in\break Table 8.

The consistency or approximate agreement of our result of
$\Delta T = 0.46 \simeq 0.5 K$ in Table 8, and the result of
$\overline{\Delta T} \cong 0.6 K$ in Table 9 requires additional
comments. The agreement strengthens the validity of our result
based on what may be called a single-layer (surface) model,
particularly evident from the equivalence of the atmosphere and
the $CO_2$-air pipe running at surface temperature and surface
pressure. On the other hand, we conclude that much of the
involved details of various flux terms of a two-layer model
can apparently be summarized by our simple radiative equilibrium
assumption for $\Delta T$ in (\ref{3.2}). For a supplementary discussion on $\Delta T$ we refer to the last part of Appendix B.

\section{On Observationally Testing the Radiative Transfer Results}
\renewcommand{\theequation}{\arabic{section}.\arabic{equation}}
\setcounter{equation}{0}

The experimental tests of predictions on the climate sensitivity in general require decadal or even centennial time scales. In pioneering papers by Lindzen and Choi \cite{Lindzen_Choi} it was pointed out that data on fluctuations of the earth's (sea-)surface temperature at short time scales, once combined with satellite-based radiation-flux measurements \cite{Barkstrom,Wielicki}, can be employed to gain important empirical information on the earth's climate sensitivity. In response to immediate reactions in the literature \cite{Chung,Murphy,Trenberth_Fasullo} on their results \cite{Lindzen_Choi}, an additional more expanded and refined investigation was presented by Lindzen and Choi 
\cite{Lindzen_Choi_new}. Compare also ref. \cite{Bates} for a more recent related investigation.

In what follows, we shall point out that the results from combining the data of sea-surface temperature fluctuations with data on radiation-flux intensities from the ERBE \cite{Barkstrom} and CERES \cite{Wielicki} experiments empirically support our results on $\Delta T$ from the radiative-transfer analysis given in Sections 2 to 4. We restrict ourselves to elaborating on the basic underlying assumptions and the final conclusion from them. Entering a discussion on the technical details involved in the analysis of the experimental data is far beyond our expertise.

In the radiative-transfer calculations in Sections 2 and 3, we proceeded in two distinct steps. In a first step, we treated a dry atmosphere (zero humidity), in the second step the realistic case of non-zero humidity was investigated. It will be useful to introduce a notation referring to these two cases explicitly. Specializing (\ref{3.2}) to the cases of zero and non-zero humidity, we define
\be
\Delta T^{(0)} = \frac{\Delta^{(0)}_{2 \times CO_2}}{\left(
\frac{dS}{dT}\right)_T} \equiv G (T) \Delta^{(0)}_{2 \times CO_2}, \label{5.1}
\ee
and
\be
\Delta T^{(H)} = \frac{\Delta^{(H)}_{2 \times CO_2}}{\left(
\frac{dS}{dT}\right)_T} \equiv G (T) \Delta^{(H)}_{2 \times CO_2}, \label{5.2}
\ee
related to zero and non-zero humidity, respectively. We recall the numerical values of $G = G (T = 293 K) = \frac{293}{4 \times 418} = 0.175 KW^{-1} m^2$, as well as $\Delta^{(0)}_{2 \times C0_2} = 6.09 Wm^{-2}$ and $\Delta^{(H)}_{2 \times C0_2} = 2.6 Wm^{-2}$ given in Section 3.

Identically rewriting $\Delta T^{(H)}$ as
\be
\Delta T^{(H)} = \Delta T^{(0)} + \Delta T^{(H)} - \Delta T^{(0)} \equiv \Delta T^{(0)} + f \Delta T^{(H)},
\label{5.3}
\ee
we find that the commonly employed so-called feedback factor $f$,
with (\ref{5.1}) and (\ref{5.2}), is given by
\be
f = 1 - \frac{\Delta T^{(0)}}{\Delta T^{(H)}} = 1 -
\frac{\Delta^{(0)}_{2 \times CO_2}}{\Delta^{(H)}_{2 \times CO_2}},
\label{5.4}
\ee
or equivalently,
\be
\Delta T^{(H)} = \frac{\Delta T^{(0)}}{1-f}.
\label{5.5}
\ee
Numerically, from (\ref{5.4}) and Table 8, we find the value of 
$f$ from the radiative-transfer calculation, henceforth called the theoretical value of $f = f_{th}$, to be given by
\be
f_{th} = 1 - \frac{6.09}{2.6} = - 1.34.
\label{5.6}
\ee
The negative feedback factor, upon insertion into (\ref{5.5}), thus identically reproduces 
\be
\Delta T^{(H)} = \frac{\Delta T^{(0)}}{2.34} = \frac{1.07}{2.34} = 0.46 K < 1.07 K
\label{5.7}
\ee
from Table 8.

Replacing $\Delta^{(H)}_{2 \times CO_2}$ in (\ref{5.4}) by
$\Delta T^{(H)}$, by employing (\ref{5.2}), the factor $f$ may be rewritten as
\be
f = 1 - G (T) \frac{\Delta^{(0)}_{2 \times CO_2}}{\Delta T^{(H)}}.
\label{5.8}
\ee
According to (\ref{5.8}), the factor $f$ is determined by dividing the radiation increase $\Delta^{(0)}_{2 \times CO_2}$ resulting from $CO_2$ doubling at zero humidity by the surface temperature increase $\Delta T^{(H)}$ necessary for equilibrium of the earth's atmosphere for $CO_2$ doubling in the case of non-zero humidity.

Identifying $\Delta^{(0)}_{2 \times CO_2} / \Delta T^{(H)}$ in
(\ref{5.8}) with the experimentally measured slope $(\Delta~{\rm Flux}/\Delta T)_{exp}$ of the outgoing radiation at the TOA measured as a function of the sea-surface temperature,
\be
\frac{\Delta^{(0)}_{2 \times CO_2}}{\Delta T^{(H)}} =
\left( \frac{\Delta ~ Flux}{\Delta T} \right)_{exp},
\label{5.9}
\ee
we find that the factor $f$ can be represented by 
\be
f \equiv f_{exp} = 1 - G(T) \left( \frac{\Delta~Flux}{\Delta T} \right)_{exp}.
\label{5.10}
\ee
Relation (\ref{5.10}) is due to Lindzen and Choi \cite{Lindzen_Choi_new} \footnote{Compare equations (5) and (6) in ref. \cite{Lindzen_Choi_new}. Equation (6) contains an additional factor $c$ via $f \to cf$ introduced for sharing of tropical feedbacks over the globe. The value of $c$ depends on how the slope $\Delta~Flux / \Delta T$ is extracted from measurements. A value of $c = 2$ is used in ref. \cite{Lindzen_Choi_new}. For a global value of $\Delta~Flux / \Delta T$ one has $c \cong 1$.}
It allows one to deduce the climate sensitivity for non-zero humidity, $\Delta T^{(H)}$, see (\ref{5.5}), from the experimentally determined radiation flux by inserting $f = f_{exp}$ according to (\ref{5.10}), and the value of $\Delta T = \Delta
T^{(0)} \simeq 1.1~K$ for $CO_2$ doubling at zero humidity.

Using satellite measurements of the ERBE \cite{Barkstrom} and CERES \cite{Wielicki} collaborations, the detailed analysis in ref. \cite{Lindzen_Choi_new} led to a value of
\be
\left( \frac{\Delta~Flux}{\Delta T} \right)_{exp} = 6.9 \pm
1.8 W m^{-2} K^{-1}
\label{5.11}
\ee
(see Table 2 in ref. \cite{Lindzen_Choi_new}). Inserting this value into (\ref{5.10}), we find
\be
f_{exp} \cong 1 - 0.175 \times 6.9 \cong - 0.21,
\label{5.12}
\ee
with $1 - f_{exp} \cong 1.21$, and $\Delta T^{(0)} \cong 1.07 K$, from Table 8, from (\ref{5.5}), we obtain
\be
\Delta T_{exp} = \frac{1.07}{1.21} \simeq 0.9 K.
\label{5.13}
\ee
Introducing the sharing factor $c=2$ from ref. \cite{Lindzen_Choi_new}, compare footnote g, we have
$f_{exp} \cong - 0.1$ and $\Delta T_{exp}$ becomes
\be
\Delta T_{exp} = \frac{1.07}{1.1} \simeq 1.0 K.
\label{5.14}
\ee

Lindzen and Choi, \cite{Lindzen_Choi_new}, employing $c = 2$,
find $f_{exp} = -0.5$ and consequently $\Delta T_{exp} \cong 0.7 K$ (confidence interval $0.5 K$ to $1.3 K$ at 99 \% confidence level).\footnote{The difference between our result of $f_{exp} = - 0.1$ for $c = 2$ and $\Delta T_{exp} \simeq 1 K$, and the result from ref. \cite{Lindzen_Choi_new}, namely $f_{exp} \simeq - 0.5$ for $c = 2$ and $\Delta T_{exp} \simeq 0.7 K$, is due to the
difference between our value of $G = G (T = 293 K) \cong 0.175 KW^{-1} m^2$ and $G (T = 255 K) \cong 0.3$ used in ref.
\cite{Lindzen_Choi_new}, or, equivalently $C^{(0)}_{2 \times CO_2} = 6.09 Wm^{-2}$ versus $C^{(0)}_{2 \times CO_2} = \Delta Q = 3.7 Wm^{-2}$ implying $T^{(0)} \simeq 1.1~K$ in both cases.}

In view of the complexity of the subject matter, we conclude that there is satisfactory agreement between the theoretical result of
\be
\Delta T_{th} = 0.46 K \simeq 0.5 K
\label{5.15}
\ee
from Table 8,\footnote{Compare also the variation of the theoretical results in Table 9 under various atmospheric conditions.} and $T_{exp} \simeq 0.9 K$ to $1.0 K$ empirically obtained upon identifying the ratio of $\Delta^{(0)}_{2 \times CO_2}/\Delta T^{(H)}$ with the satellite-based radiation flux measurements, $\Delta~ Flux / \Delta T$. In fact, we conclude
that the results from the satellite-based radiation-flux measurement support the results on the climate sensitivity from the detailed and reliable radiation-transfer calculations.

As pointed out in refs. \cite{Lindzen_Choi} and \cite{Lindzen_Choi_new}, there is definite disagreement between the results from the radiation-flux measurements of $f_{exp} < 0$, and the predictions of a large number of atmospheric models that arrive at a positive feedback factor $f > 0$ and an enhanced value of the climate sensitivity of $\Delta T ({\rm atm.~models}) \gg 1 K$.

We add a reference to a recent paper \cite{Kauppinen} that responds to the claim \cite{IPCC} of empirical support for a significant climate sensitivity, $\Delta T \gg 1K$, due to $CO_2$ doubling, based on the analysis of the observed changes of
$\Delta T$ in the 25-year period from 1983 to 2008. It is pointed out \cite{Kauppinen} that the climate models employed \cite{IPCC} for this analysis fail to correctly incorporate the important influence of the low-cloud-cover
contribution on $\Delta T$ in the period under investigation. Correctly incorporating the cloud cover effect yields consistency with $\Delta T < 1 K$ \cite{Kauppinen} in disagreement with the results of $\Delta T \gg 1 K$ \cite{IPCC}.

\section{Conclusions}

It has been the aim of this paper to estimate the increase in
temperature $\Delta T$ (``climate sensitivity'') of the surface
of the earth due to a doubling of the $CO_2$ concentration in the
atmosphere.
The estimate is obtained in a concise and transparent
manner without oversimplification. All necessary steps are
explicitly elaborated upon.

The basic assumption of associating a uniform constant temperature $T$
with the surface of the earth, and a black-body long-wave infrared
radiation $S(T)$, is by no means trivial, implicitly or explicitly, however,
common to 
main-stream investigations on this matter. Our results are based on a new
radiative-transfer evaluation, the details being presented in Appendix A.
The absorption of the
atmosphere in the $CO_2$ spectral range can be, and is reliably determined,
and leads to an approximately constant value beyond an altitude of about
5 km, or a length of the horizontal $CO_2$-air pipe of about 3 km at surface
temperature and pressure.

Assuming restoration of equilibrium upon
doubling of the $CO_2$ concentration by an associated increase of the
temperature then implies a definite estimate of the increase of the surface
temperature $\Delta T$, given by $\Delta T \cong 0.5~^0C$ 
(compare Sections 3 and 4).

In terms of the widely employed feedback parameter $f$, the result of $\Delta T \cong 0.5^0 C$ corresponds to a negative feedback of $f < 0$. This result is empirically supported by satellite-based measurements of short-time fluctuations of the outgoing radiation flux at the TOA as a function of (sea-)surface temperature. A consistent picture emerges by combining theoretical radiation-transfer results with radiation-flux measurements (compare Section 5). This picture disagrees with an abundant number of predictions from climate models that imply positive feedbacks, $f > 0$.

The quantitative result of $\Delta T \simeq 0.5$ to $0.6~ ^0C$ valid for
the drastic increase of doubling of the $CO_2$ content in air from
380 ppm to 760 ppm to be related to one century, confirms that the effect of
an anthropogenic $CO_2$ increase
on the climate on earth is fairly negligible. This conclusion is in strong
contrast to the values of $\Delta T \sim 1.5 - 4.5~ ^0C$ quoted in the
2013 IPCC report\cite{IPCC}.
The published results on $\Delta T$ fill an even larger interval between
$\Delta T \simeq 0.4 ^0C$ to $\Delta T \simeq 8 ^0C$.
There is a systematic
tendency of the results on $\Delta T$ published between the years 2000 to 2018
to decrease\cite{Gervais_ESR} with increasing publication date,
the results coming
closer to our result of $\Delta T \simeq 0.5~ ^0C$.\footnote{Compare e.g. Fig. 1 in ref. \cite{Gervais_ESR} and the associated list of tens of references.}

We found a concise and brief derivation of the earth's greenhouse effect
(compare Appendix B, in particular Fig. B1). The greenhouse effect is
explicitly derived. It is understood as a shift of the radiative
equilibrium from the earth's surface to an atmospheric level above
approximately five kilometers.

\section*{Acknowledgements}

The author thanks Konrad Kleinknecht for discussions that initiated
the present investigation. The author thanks Peng-Sheng Wei for a
useful correspondence. Particular thanks go to Roy Clark for providing
the radiative transfer results in Appendix A, for much help in
tracing the relevant literature on the subject matter, as well as a critical
reading of the present paper.

\newpage

\appendix{$\mathbf CO_2$ Absorbance 568 to 770 {\boldmath${\rm cm}^{-1}$},
7 km path, 296 K, 300, 600 ppm}

The configurations used for the pipe and atmospheric absorption calculations
are shown in 
Figures A1a and A1b.  

\begin{figure}[h!] 
\centerline{\psfig{file=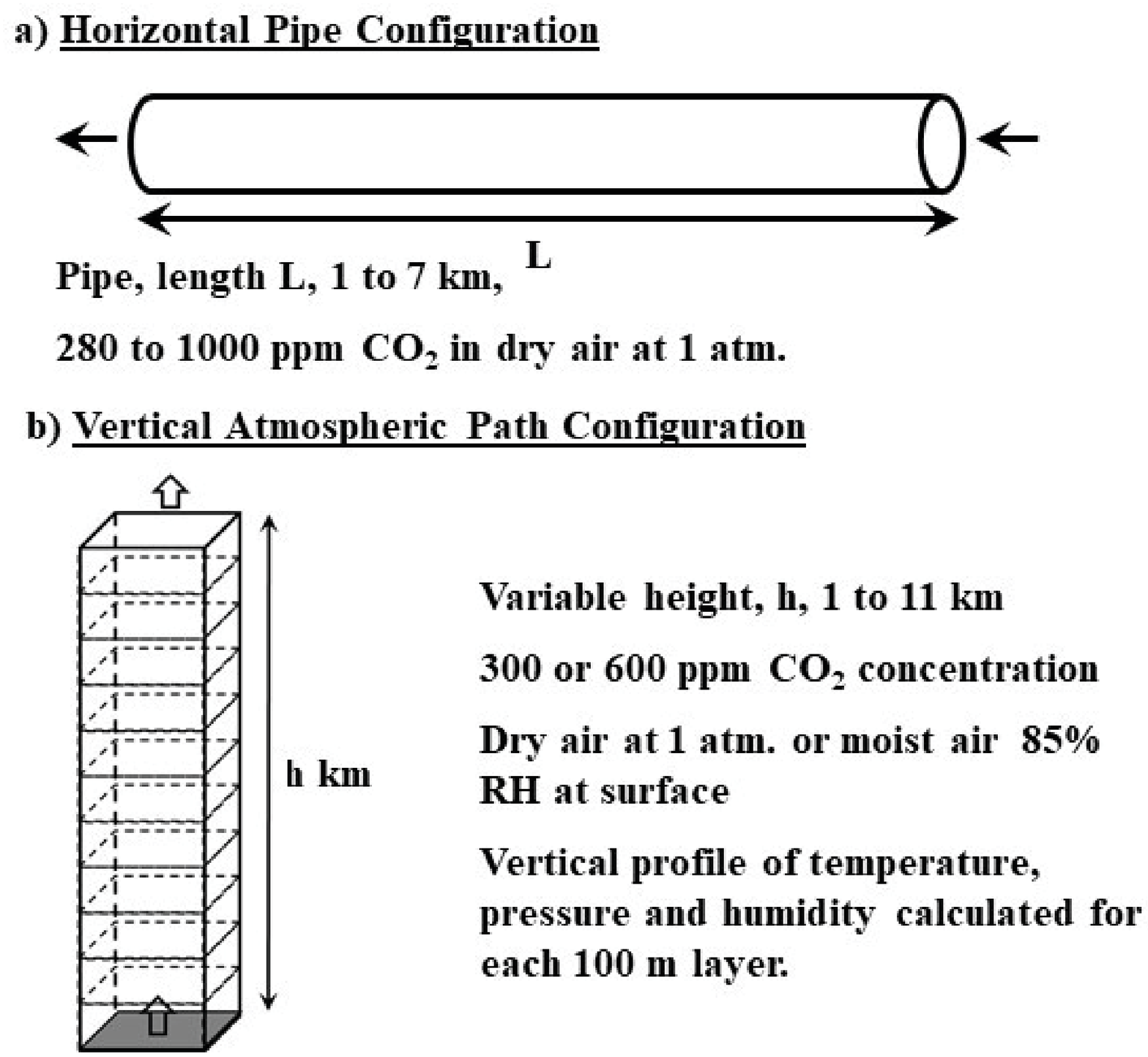,width=2.5in}}
\vspace*{8pt}
\caption{Configurations used for the pipe and atmospheric absorption
calculations.}
\end{figure}

For the `pipe' configuration, a variable horizontal
path length L, from 1 to 
7 km long was used.  The $CO_2$ concentration was varied from 280 to 1000 ppm.
A total pressure 
of 1 atmosphere dry air at 293 K was used.  

For the atmospheric absorption
calculation, a vertical 
path, h, from 1 to 11 km was used.  The surface temperature was set
to 293 K at 1 atmosphere 
pressure.  Two sets of calculations were performed.  The first used dry
air with a lapse rate of - 6.5 K ${\rm km}^{-1}$.  The second used a
surface relative humidity of 85 \% with a lapse rate of - 6.5 K
${\rm km}^{-1}$ up to the saturation level.  Above the saturation level,
the meteorological formula for the 
saturated lapse rate is used\cite{Tsonis}.  The lapse rates used are plotted
below in Fig. A6.  
The vertical atmospheric profile was calculated using a spatial resolution
of 100 m.

The high resolution CO2 absorbance spectra were calculated using
spectral data from the 
HITRAN database\cite{HITRAN}, $^{12}C^{16}O_2$ lines only with linestrengths $ >$ 1e-23.
The spectral resolution was 0.01 ${\rm cm}^{-1}$.  The temperature was
set to 293 K.  For the `pipe' calculations, the large spectral data 
files were divided into seven spectral ranges with the absorption
coefficient calculated for each 
spectral range.  The calculations were performed using Excel workbooks
with each spectral 
range in a separate workbook.  The weighted averages over the spectral
ranges 568 to 770 ${\rm cm}^{-1}$ 
and 600 to 750 ${\rm cm}^{-1}$ were then calculated. The `pipe' calculation
is illustrated in Table A1.
Path lengths of 1, 2, 3, 5
and 7 km were used.  The 
results are given in Tables A2 and A3.
Plots of the data are shown in Figures A2 and A3.
\newpage

\begin{table}[h!]   
\tbl{The weighted averages of the seven spectral ranges are combined
to give the band average absorption.\break
R. Clark, April $28^{th}$ 2019}
{\begin{tabular}{@{}cccc@{}} \Hline 
\\[-1.8ex] 
{\bf Range {\boldmath${\rm cm}^{-1}$}} & {\bf Start/End} & {\bf 300 ppm} & {\bf 600 ppm} \\
\Hline
32 & 568 to 600 ${\rm cm}^{-1}$ & 0.51 & 0.64 \\
20 & 600 to 620 ${\rm cm}^{-1}$ & 0.94 & 0.99 \\
20 & 620 to 640 ${\rm cm}^{-1}$ & 0.99 & 1.00 \\
60 & 640 to 700 ${\rm cm}^{-1}$ & 1.00 & 1.00 \\
25 & 700 to 725 ${\rm cm}^{-1}$ & 0.99 & 1.00 \\
25 & 725 to 750 ${\rm cm}^{-1}$ & 0.89 & 0.97 \\
20 & 750 to 770 ${\rm cm}^{-1}$ & 0.39 & 0.53 \\
{\bf Band Average} & & & \\
\Hline
202 & 568 to 770 $cm^{-1}$ & 0.84 & 0.89 \\
& & & \\
& {\bf Absn coeff -ln(1-A)} & {\bf 1.83} & {\bf 2.21} \\
& {\bf a ({\boldmath${\rm atm}^{-1}~ {\rm m}^{-1}$})} & {\bf 0.870}
& {\bf 0.527} \\
\Hline
\end{tabular}}
\label{tabA1}
\end{table}

\begin{table}[h!]   
\tbl{`Pipe' absorption for selected path lengths from 1 to 7 km,
$CO_2$ concentrations of 280 to 1000 ppm, 568 to 769 
${\rm cm}^{-1}$ spectral range, 293 K.}
{\begin{tabular}{@{}lccccccc@{}} \Hline 
\\[-1.8ex] 
ppm & 280 & 300 & 380 & 400 & 600 & 760 & 1000 \\
\Hline
1 km & 0.62 & 0.63 & 0.66 & 0.66 & 0.71 & 0.73 & 0.76 \\
2 km & 0.70 & 0.71 & 0.73 & 0.74 & 0.78 & 0.81 & 0.83 \\
3 km & 0.75 & 0.75 & 0.78 & 0.78 & 0.82 & 0.84 & 0.87 \\
5 km & 0.80 & 0.81 & 0.83 & 0.83 & 0.87 & 0.88 & 0.90 \\
7 km & 0.83 & 0.84 & 0.86 & 0.86 & 0.89 & 0.90 & 0.92 \\
\Hline
\end{tabular}}
\label{tabA2}
\end{table}

\begin{figure}[h!] 
\centerline{\psfig{file=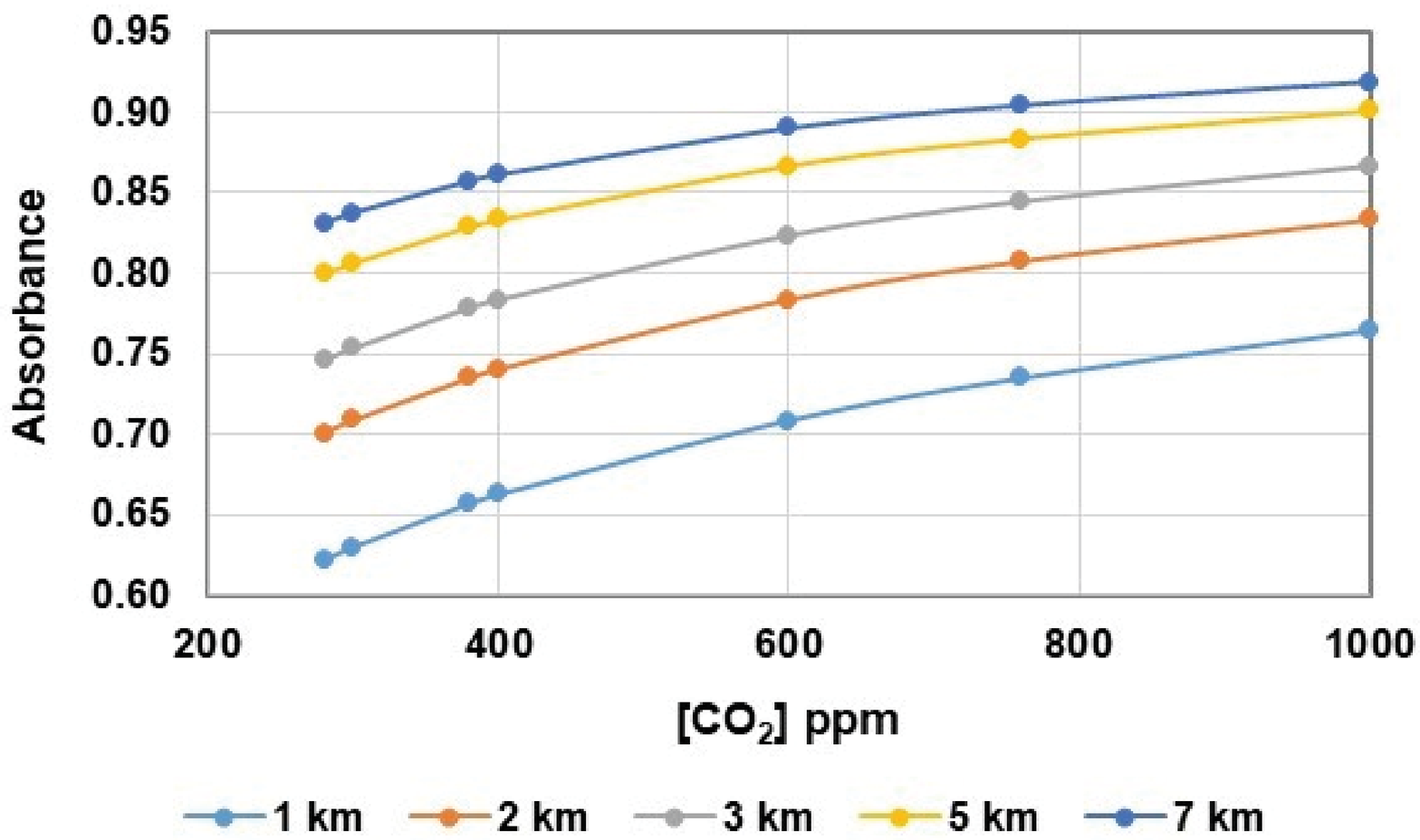,width=2.5in}}
\vspace*{-1.2cm}
\caption{Data plotted from Table A2.}
\end{figure}
\newpage
\begin{table}[h!]   
\tbl{`Pipe' Absorption for selected path lengths and concentrations,
1 to 7 km and 280 to 1000 ppm, 600 to 750 ${\rm cm}^{-1}$ spectral range}
{\begin{tabular}{@{}lccccccc@{}} \Hline 
\\[-1.8ex] 
ppm & 280 & 300 & 380 & 400 & 600 & 760 & 1000\\
\Hline
1 km & 0.78 & 0.79 & 0.82 & 0.83 & 0.87 & 0.89 & 0.92 \\
2 km & 0.86 & 0.87 & 0.89 & 0.90 & 0.93 & 0.95 & 0.97 \\
3 km & 0.90 & 0.91 & 0.93 & 0.93 & 0.96 & 0.97 & 0.98 \\
5 km & 0.94 & 0.95 & 0.96 & 0.97 & 0.98 & 0.99 & 0.99 \\
7 km & 0.96 & 0.97 & 0.98 & 0.98 & 0.99 & 1.00 & 1.00 \\
\Hline
\end{tabular}}
\label{tabA3}
\end{table}

\begin{figure}[h!]
\centerline{\psfig{file=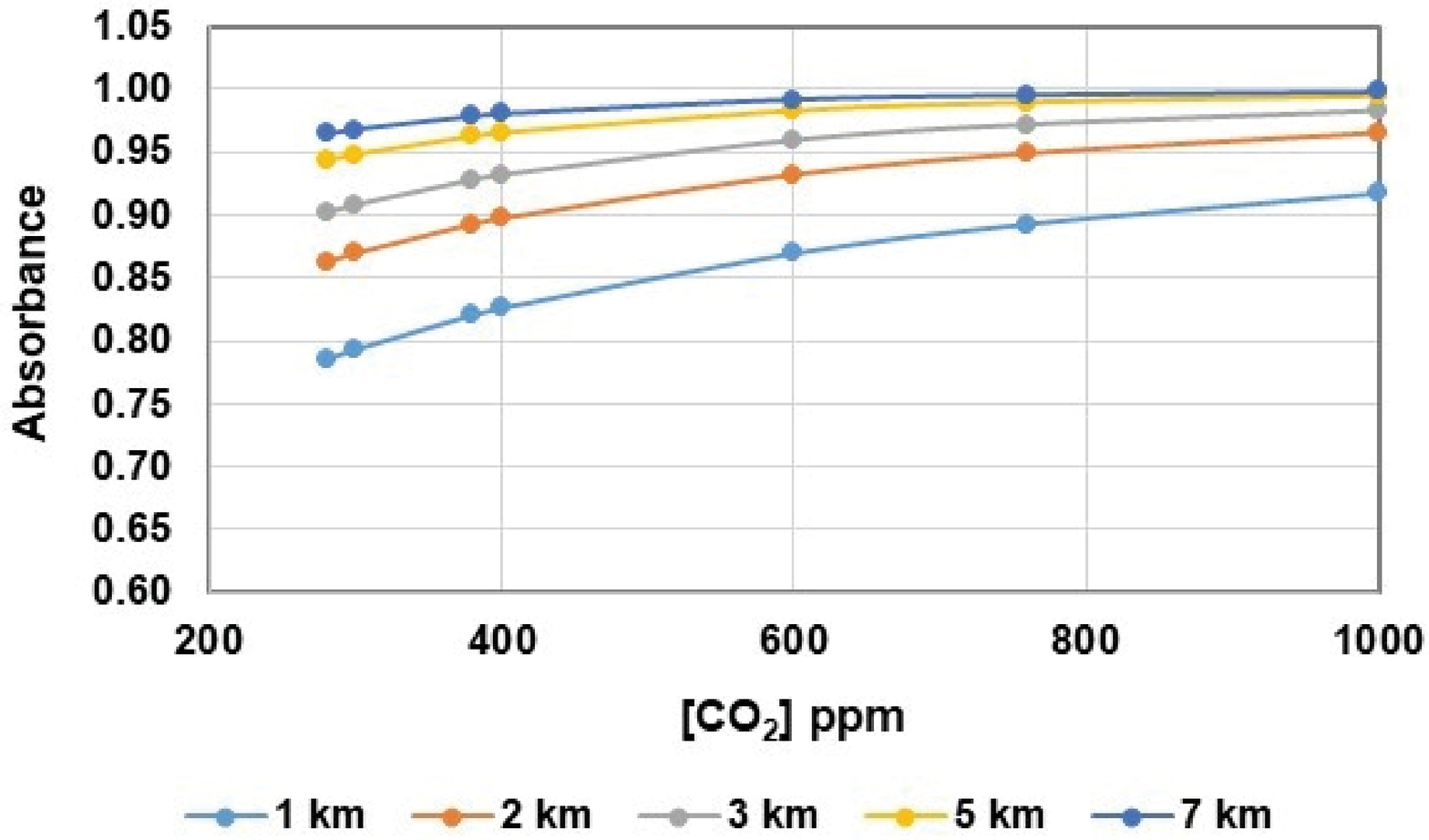,width=2.5in}}
\vspace*{-1.2cm}
\caption{Data plotted from Table A3.}
\end{figure}

The atmospheric radiative transfer calculations were performed using
the same $CO_2$ lines that were used for the `pipe' calculations with
$^1H_2^{16}O$ lines with linestrengths $>$ 1e-23 added for the 
85 \% RH cases.  The calculations were performed using Excel VBA macros
at a spectral resolution of 0.01 ${\rm cm}^{-1}$.  However, the data were
binned into 1 ${\rm cm}^{-1}$ intervals for plotting.  The 
conditions of temperature and $CO_2$ concentration and the spectral ranges
were the same as for the 
`pipe' data.  The $H_2O$ concentration was set to a relative humidity
of 85 \% at 293 K for the first 100 m layer.  The vertical 
path length was varied from 1 to 11 km in 1 km steps.
The results are given in Tables A4 and 
A5.  Plots of the data are shown in Figures A4 and A5.

\begin{table}[h!]   
\tbl{Atmospheric absorption data for the spectral range from
568 to 769 ${\rm cm}^{-1}$ and from 600 to 750 ${\rm cm}^{-1}$.  The vertical heights from the surface were varied
from 1 to 11 km.}
{\begin{tabular}{@{}ccccc@{}} \Hline 
\\[-1.8ex] 
 & 568 to 769 ${\rm cm}^{-1}$ & 568 to 769 ${\rm cm}^{-1}$
 & 600 to 750 ${\rm cm}^{-1}$ & 600 to 750 ${\rm cm}^{-1}$\\
\Hline
Altitude km & 300 ppm & 600 ppm & 300 ppm & 600 ppm \\
\Hline
1 & 0.618 & 0.696 & 0.777 & 0.856 \\
2 & 0.682 & 0.758 & 0.842 & 0.910 \\
3 & 0.713 & 0.786 & 0.872 & 0.933 \\
4 & 0.731 & 0.802 & 0.888 & 0.945 \\
5 & 0.742 & 0.811 & 0.897 & 0.952 \\
6 & 0.749 & 0.817 & 0.904 & 0.956 \\
7 & 0.754 & 0.821 & 0.908 & 0.958 \\
8 & 0.757 & 0.823 & 0.910 & 0.960 \\
9 & 0.759 & 0.825 & 0.912 & 0.961 \\
10 & 0.761 & 0.826 & 0.913 & 0.962 \\
11 & 0.761 & 0.827 & 0.914 & 0.962 \\
\Hline
\end{tabular}}
\label{tabA4}
\end{table}

\begin{figure}[h!] 
\centerline{\psfig{file=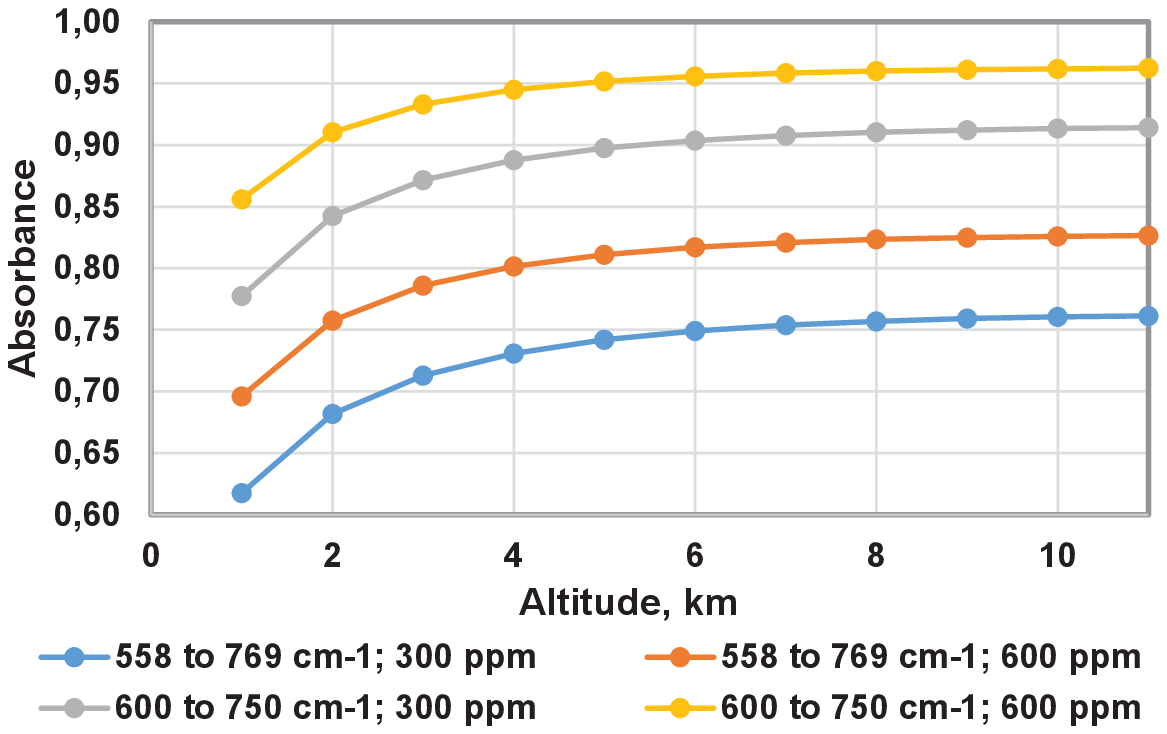,width=2.5in}}
\vspace*{8pt}
\caption{Data plotted from Table A4.}
\end{figure}

\begin{table}[h!]   
\tbl{Atmospheric absorption data for the spectral range from 558 to 769 ${\rm cm}^{-1}$ and from
600 to 750 ${\rm cm}^{-1}$.  The vertical heights from the surface were varied from 1 to 11 km.}
{\begin{tabular}{@{}ccccc@{}} \Hline 
\\[-1.8ex] 
& 558 to 769 ${\rm cm}^{-1}$ & 558 to 769 ${\rm cm}^{-1}$
 & 600 to 750 ${\rm cm}^{-1}$ & 600 to 750 ${\rm cm}^{-1}$\\
\Hline
Altitude km & 300 ppm, 85 RH & 600 ppm, 85 RH & 300 ppm, 85 RH
& 600 ppm, 85 RH \\
\Hline
1 & 0.799 & 0.843 & 0.856 & 0.904 \\
2 & 0.854 & 0.892 & 0.906 & 0.945 \\
3 & 0.876 & 0.910 & 0.927 & 0.961 \\
4 & 0.887 & 0.919 & 0.937 & 0.968 \\
5 & 0.894 & 0.924 & 0.943 & 0.973 \\
6 & 0.897 & 0.927 & 0.947 & 0.975 \\
7 & 0.900 & 0.929 & 0.949 & 0.977 \\
8 & 0.901 & 0.930 & 0.951 & 0.977 \\
9 & 0.902 & 0.931 & 0.952 & 0.978 \\
10 & 0.903 & 0.931 & 0.952 & 0.978 \\
11 & 0.903 & 0.929 & 0.953 & 0.977 \\
\Hline
\end{tabular}}
\label{tabA5}
\end{table}
\begin{figure}[h!] 
\centerline{\psfig{file=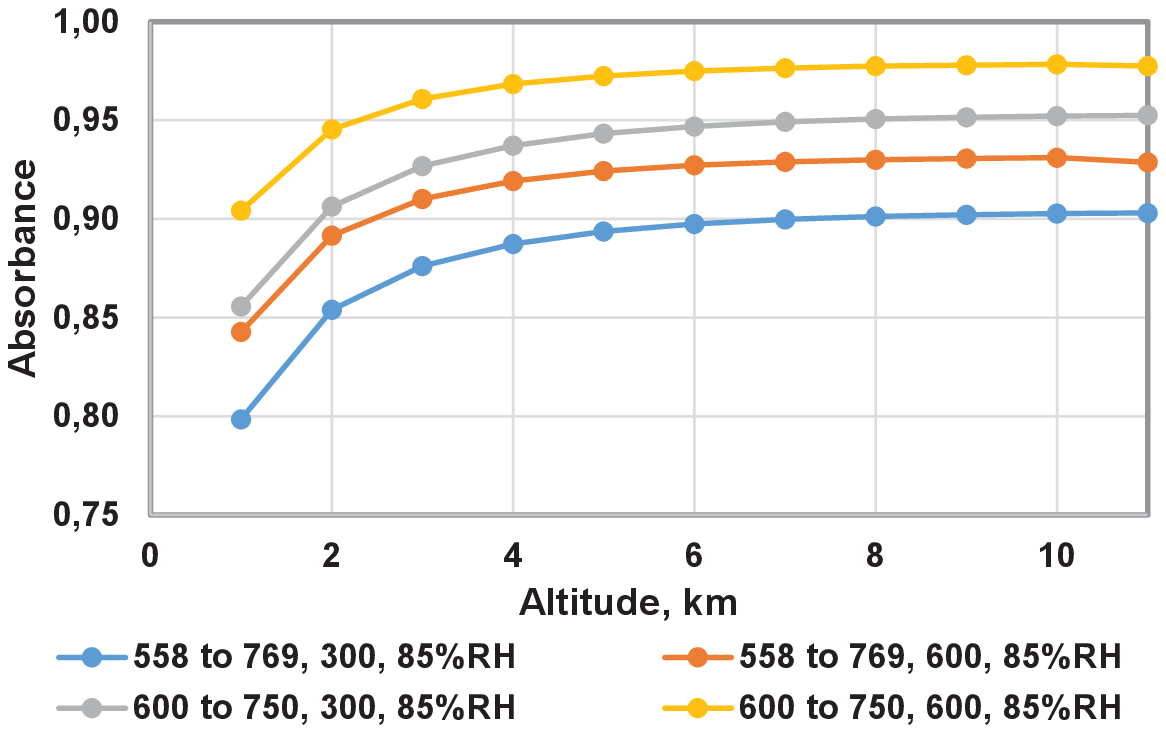,width=2.5in}}
\vspace*{-0.1cm}
\caption{Data plotted from Table A5.}
\end{figure}


\begin{figure}[h] 
\centerline{\psfig{file=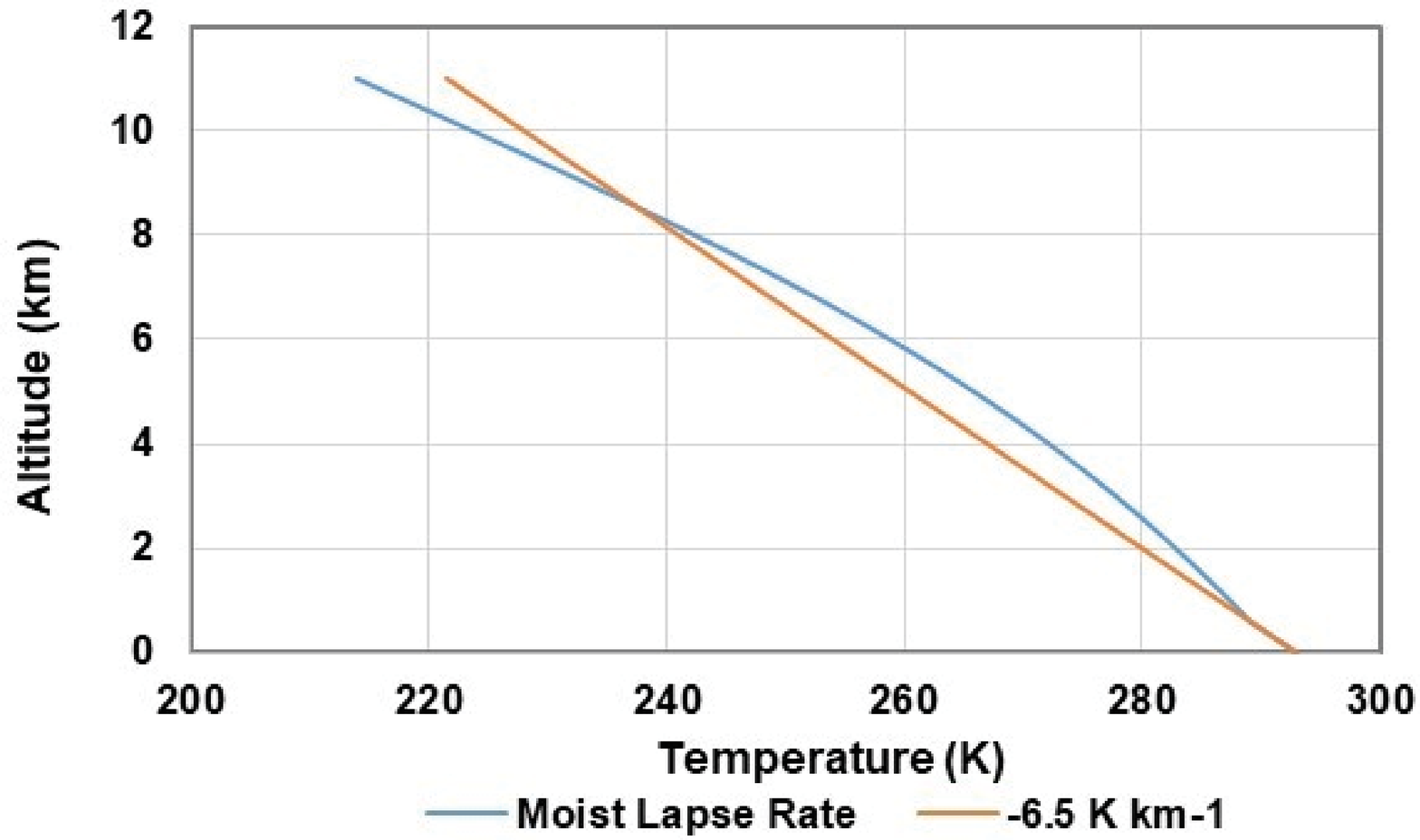,width=2.5in}}
\vspace*{-1.2cm}
\caption{The lapse rates used in the atmospheric absorption calculations,
from $^{23}$.}
\end{figure}
\newpage

\appendix{A Concise Representation of the Radiative ``Greenhouse Effect'' in
the Atmosphere}
The surface of the earth receives a net radiation from the sun, $S_0$,of
approximately $S_0 \simeq 240 Wm^{-2}$, see footnote b
in the main text. The infrared radiation from the surface of the earth,
the LWIR,
contains a component, $S_W$, restricted in its spectral range to the
so-called ``atmospheric window'', and a second component, $S_{up}$.
The second component consists of a small part, $f^\prime~S_{up}~
(f^\prime \ll 1)$, emitted to space, and a larger fraction, $(1-f^\prime)~
S_{up}$ that is absorbed by the so-called
``greenhouse gases'', essentially $H_2O$ and $CO_2$, in the atmosphere,
and subsequently re-emitted, partly to space, $A_{up}$, and partly down
to the surface, $A_{down}$, compare Fig. B1.

The ``transmittance parameter'' $f^\prime$, with $0 \le f^\prime \le 1$,
realistic
values being restricted to $f^\prime \ll 1$, quantifies the transmittance
of the earth's surface radiation in the spectral range
outside the atmospheric window. Compare the incomplete saturation of
the absorption in Fig. 4a at altitudes larger than (approximately) 9 km.

\begin{figure}[h] 
\centerline{\psfig{file=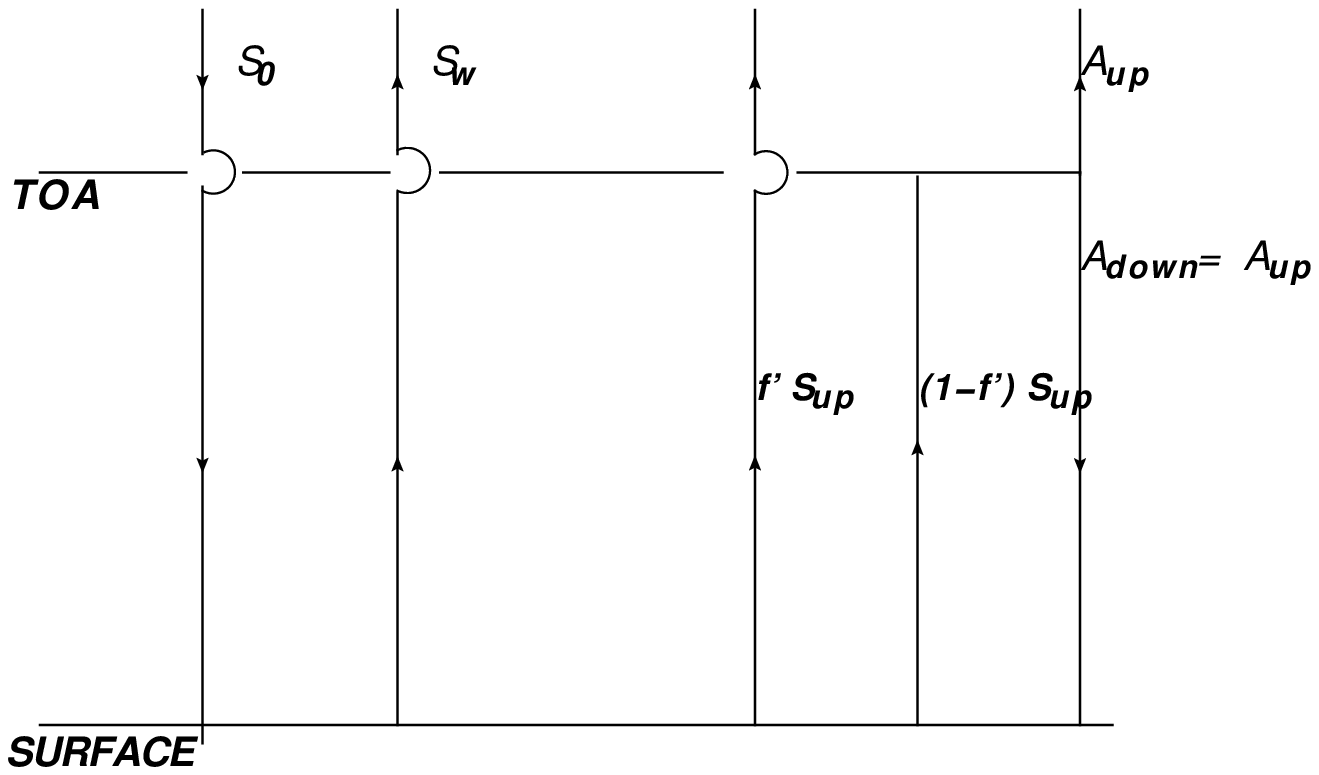,width=3in}}
\vspace*{8pt}
\caption{Schematic representation of the net radiation from the sun,
$S_0 \simeq 240 Wm^{-2}$, the infrared radiation to space through the
atmospheric window, $S_W$, the infrared radiation from the surface
to space $f^\prime S_{up}$, and to the atmosphere $(1-f^\prime)~S_{up}$,
where $f^\prime \ll 1$, and the radiation from the atmosphere to
space $A_{up}$ and to the surface, $A_{down}$.}
\end{figure}

Global radiative equilibrium of the earth is due to an equality between
the intensity of the short-wave radiation received from the sun, $S_0$,
and the infrared radiation sent out to space, i.e.
\be
S_0 = S_W + f^\prime S_{up} + A_{up}.
\label{A2.1}
\ee

Equilibrium at the upper atmosphere (the top of the atmosphere, TOA) requires
\be
(1 - f^\prime) S_{up} = A_{up} + A_{down} = 2 A_{up},
\label{A2.2}
\ee
where equality of the upward and the downward emission from the greenhouse
gases has been employed in the second step. From (\ref{A2.1}) and
(\ref{A2.2}), we deduce the radiation from the surface, $S_{up}$, given by
\be
S_{up} = \frac{2}{1 + f^\prime} (S_0 - S_W),
\label{A2.3}
\ee
as well as the radiation from the TOA,
\be
A_{up} = A_{down} = \frac{1 - f^\prime}{1+f^\prime} (S_0 - S_W).
\label{A2.4}
\ee

Both, $S_{up}$ and $A_{up}$ are expressed in terms of $S_0 - S_W$, that
is the radiation from the sun upon reduction by the radiation, $S_W$,
that is associated with the transmission of radiation through the
atmospheric window.

For given $S_0$, from (\ref{A2.3}), we find that the total emission
by the surface, $S_{tot, surf} (S_W, f^\prime)$, is given by
\be
S_{tot,surf} (S_W, f^\prime) = S_W + S_{up} = S_0 \left( \frac{2}{1+f^\prime}
- \frac{S_W}{S_0} \frac{1 - f^\prime}{1+f^\prime} \right) \equiv
S_0 c \left( \frac{S_W}{S_0}, f^\prime \right).
\label{A2.5}
\ee

Employing (\ref{A2.3}) and (\ref{A2.4}), one may explicitly check
the validity of equilibrium at the surface, 
\be
S_{tot,surf} = S_W + S_{up} = S_0 + A_{down} = S_0 + A_{up}.
\label{A2.6}
\ee

The radiation to space at the TOA,
\be
S_{tot, TOA} (S_W, f^\prime) = S_W + f^\prime S_{up} + A_{up},
\label{A2.7}
\ee
upon substitution of the results for $S_{up}$ and for $A_{up}$ from
(\ref{A2.3}) and (\ref{A2.4}) becomes
\be
S_{tot, TOA} (S_W, f^\prime) = S_0.
\label{A2.8}
\ee
Global radiative equilibrium (\ref{A2.1}) is realized by the equality
of the radiation intensities emitted from the TOA, $S_{tot, TOA}$,
and received from the sun, $S_0$.

It will be illuminating to numerically evaluate our main result, the
expression for the surface radiation (\ref{A2.5}), for several
relevant values of $S_W$ and $f^\prime$ for fixed $S_0 = 240 Wm^{-2}$.

Upon assuming a Planck black-body radiation spectrum for the surface
radiation (\ref{A2.5}),
\be
S_{tot, surf} (S_W, f^\prime) = \sigma \left( T_{surf} (S_W, f^\prime)
\right)^4,
\label{A2.9}
\ee
where $\sigma \cong 5.670 \times 10^{-8} Wm^{-2} K^{-4}$, we obtain the
associated surface temperature,
\bqa
T_{surf}(S_W, f^\prime) & = \left( \frac{S_0}{\sigma}\right)^{\frac{1}{4}}
\left( \frac{2}{1 + f^\prime} - \frac{S_W}{S_0} \frac{1-f^\prime}{1+f^\prime}
\right)^\frac{1}{4} \nonumber \\
& = 255.069 \left( \frac{2}{1+f^\prime} - \frac{S_W}{S_0}
\frac{1-f^\prime}{1+f^\prime} \right)^{\frac{1}{4}} K \nonumber \\
& = 255.069 c \left(\frac{S_W}{S_0}, f^\prime\right)^{\frac{1}{4}} K.
\label{A2.10}
\eqa

Relations (\ref{A2.5}) and (\ref{A2.10}) give the equilibrium surface
radiation and surface temperature for given values of $S_0, S_W$ and
$f^\prime$.

The numerical results for $S_{tot,surf}$ from (\ref{A2.5}) and
$T_{surf}$ from (\ref{A2.10}) for a few representative values of
$S_W$ are collected in Table B1.

\begin{table}[h]   
\tbl{Numerical results for the total surface radiation
$S_{tot, surf} (S_W, f^\prime)$ given by (\ref{A2.5}), and for
the surface temperature $T_{surf} (S_W, f^\prime)$ given by
(\ref{A2.10}).}
{\begin{tabular}{@{}cccc@{}} \Hline 
\\[-1.8ex] 
$S_W [Wm^{-2}] (f^\prime)$ & $c \left( \frac{S_W}{S_0}, f^\prime \right)$
& $S_{tot, surf} [Wm^{-2}]$ & $T_{surf} [K]$ \\
\Hline
240 & 1.0 & 240 & 255 = -- $18^0$ C \\
and/or $f^\prime = 1$ & & & \\
58 & 1.62 & 389 & 288 = $15^0$ C \\
$(f^\prime = 0.1)$ & & & \\
43 & 1.67 & 401 & 290 = $17^0$ C \\
$(f^\prime = 0.1)$ & & & \\
28 & 1.72 & 413 & 292 = $19^0$ C \\
$(f^\prime = 0.1)$ & & & \\
0 & 1.82 & 436 & 296 = $23^0$ C \\
$(f^\prime = 0.1)$ & & & \\
0 & 2.0 & 480 & 303 = $30^0$ C \\
$(f^\prime = 0)$ & & & \\
\Hline
\end{tabular}}
\label{tabB1}
\end{table}

Turning to a discussion of the results in Table B1, we start with
the results for the empirical input of $S_W = 43 \pm 15 Wm^{-2}$, or
$58 \ge S_W \ge 28 Wm^{-2}$, deduced\cite{Clark2} from meteorological
measurements of the long-wave infrared radiation (LWIR). The value of
$f^\prime = 0.1$ follows from Table 6 in the main text. At the saturation
limit (for $C0_2$ content of 300 ppm) the absorption reaches a value of
$A = 0.90$. This corresponds to $(1 - f^\prime) S_{up} = 0.90 S_{up}$,
or $f^\prime = 0.1$.
The results
for the surface temperature, $T_{surf}$, from (\ref{A2.10}) given
by $288 \le T_{surf}
\le 292 K$, or $15^0$ C to $19^0$ C, are consistent with the very
elaborate theoretical analysis in refs.\cite{Trenberth} and\cite{Harde}.

The case of $S_W = S_0 = 240 Wm^{-2}$ in the first line of Table B1, according
to (\ref{A2.3}) and (\ref{A2.4}), corresponds to $S_{up} = A_{up} = 0$.
This is the hypothetical case of an atmosphere allowing for direct
total emission from the surface to space, or zero absorption, no
greenhouse gas present. Comparing $S_{tot~surf} (S_W = S_0, f^\prime)$
with (\ref{A2.8}), we find the identity
\be
S_{tot~TOA} (S_W, f^\prime) = S_{tot~surf} (S_W = S_0, f^\prime).
\label{A2.11}
\ee
The infrared radiation at the TOA, $S_{tot~TOA} (S_W, f^\prime) = S_0$,
is equal to the surface radiation of a non-absorbing
hypothetical atmosphere\footnote{Equivalently, the non-absorbing atmosphere
may be described by inserting $f^\prime = 1$ in (\ref{A2.5}), implying
$S_{tot~surf} (S_W, f^\prime = 1) = S_0$ and $A_{up} = A_{down} = 0$ according
to (\ref{A2.4}).}. The associated temperature is given by
$T_{TOA} = T_{surf} = 255 K$. The ``greenhouse effect'' of the
atmosphere containing ``greenhouse gases'' consists of shifting the
origin of the outgoing infrared radiation (responsible for equilibrium
with the radiation from the sun) from the surface to the TOA.

The hypothetical limit of $S_W = 0$ and $f^\prime = 0$ in
the last line of Table B1
describes the case of a hypothetical ``closed atmospheric window'',
total absorption of the emitted surface radiation of $S_{tot, surf}
(S_W = 0, f^\prime = 0) = 2 S_0$, implying the maximally possible
equilibrium surface temperature of $T_{surf} = 303 K = 30 ^0C$.

A comment on the consistency of our results with the 
radiation and energy budget analysis in ref.\cite{Trenberth} is
appropriate. Trenberth et al.\cite{Trenberth}, using $S_0^\prime =
239 Wm^{-2}$ as input solar radiation, find a total surface radiation,
$S^\prime_{tot~surf} = 396 Wm^{-2}$ corresponding to $T^\prime = 289 K =
16 ^0C$, consistent with our result from Table B1, $S_{tot~surf} \simeq
401 Wm^{-2}$ and $T = 17 ^0C$. The consistency is not accidental, but can
be understood as follows.

According to ref.\cite{Trenberth}, in our notation, the downward radiation
is given by $A^\prime_{down} = 333 Wm^{-2}$. Including the amount of
$78 Wm^{-2}$\cite{Trenberth} of solar radiation absorbed by the atmosphere
into the solar radiation absorbed by the surface, one effectively has
$A^{\prime \prime}_{down} = 333 - 78 = 255 Wm^{-2}$.
Thermals and evaporation amount to $97 Wm^{-2}$\cite{Trenberth}. From
our point of view they should be considered as being independent from the
radiation budget. The amount of $97~ Wm^{-2}$ quantifies the
energy conversion of an adiabatic heat engine working between the
temperatures of $T = 289 K$ on the surface, and $T = 255 K$ at the TOA.
The heat engine, working in the gravitational field of the earth,
solely continuously converts internal energy of the air at $T = 289 K$
back and forth into potential energy at $T = 255 K$. This process does not
affect the radiative balance, and accordingly, $A^\prime_{down} = 333 Wm^{-2}$
is reduced to $A^{\prime\prime\prime}_{down} = 333 - 78 - 97 = 158 Wm^{-2}$,
a value consistent with $A_{down} = 161 Wm^{-2}$ from (\ref{A2.4}).

The radiation balance obtained by Trenberth et al.\cite{Trenberth}
accordingly is consistently reduced to the simple representation
shown in Fig. C1 that contains the essential features of the
greenhouse effect.

We turn to the change of the surface radiation under doubling
of the atmospheric $CO_2$ content, and the associated change of the
surface temperature $\Delta T$ discussed in section 3 of the main
text.

For a small change of the parameter $f^\prime \to f^\prime + \Delta f^\prime$,
with $S_W = const$, from (\ref{A2.5}) and (\ref{A2.6}) we find
\be
\frac{dS_{up} (S_W, f^\prime)}{df^\prime} = \frac{dS_{tot,~surf} (S_W,f^\prime)}
{df^\prime} = - \frac{2(S_0 - S_W)}{(1+f^\prime)^2},
\label{C.12}
\ee
or
\be
\Delta f^\prime = - \frac{(1+f^\prime)^2}{2 (S_0 - S_W)} \Delta S_{up} =
- \frac{(1+f^\prime)^2}{2(S_0 - S_W)} \Delta_{2 \times CO_2},
\label{C.13}
\ee
where $\Delta S_{up} = \Delta S_{tot,~surf} = \Delta_{2 \times CO_2}$,
due to $CO_2$ doubling, was inserted in the second step. The negative
value of $\Delta f^\prime < 0$ for $\Delta_{2 \times CO_2} > 0$ implies
less direct radiation to space and, accordingly, a rise of the surface
temperature, $\Delta T > 0$.

For the change of the surface temperature $T_{surf}$, from (\ref{A2.10}), we
obtain
\be
\Delta T = \frac{dT}{df^\prime} df^\prime = - \frac{T_{surf}}{4}
\frac{2 \left( 1 - \frac{S_W}{S_0}\right)}{2 (1+f^\prime) -
\frac{S_W}{S_0} (1 - f^{\prime 2})} \Delta f^\prime =
\frac{T_{surf}}{4} \frac{\Delta_{2 \times CO_2}}{S_{tot,~surf} (S_W, f^\prime)},
\label{C.14}
\ee
where $\Delta f^\prime$ from (\ref{C.13}) was inserted in the last step.
The expression resulting from the insertion was simplified to yield
the expression for $S_{tot,~surf} (S_W, f^\prime)$ in (\ref{A2.5})
in the denominator of (\ref{C.14}).

The agreement of $\Delta T$ in (\ref{C.14}) with $\Delta T$ in
(\ref{3.2}) of the main text comes without surprise. We note however, that
the present derivation of (\ref{C.14}), relying on (\ref{A2.5}) and
(\ref{A2.10}), assures equilibrium of radiation at the earth's surface and
at the TOA for any given values of $S_0, S_W, f^\prime$, and for
$f^\prime \to f^\prime + \Delta f^\prime$ in particular. In slight
distinction from the present derivation of (\ref{C.14}), in (\ref{3.2}),
we relied on using restoration of equilibrium upon $CO_2$ doubling.
It is noteworthy, moreover, that $T_{surf}$ in (\ref{A2.10}) and
$\Delta T$ in (\ref{C.14}) is predicted from the empirically known
values of $S_0, S_W$ and $f^\prime$, while $T_{surf}$ in (\ref{3.2}) was
used as an input parameter.

\begin{table}[h]   
\tbl{Numerical results for the total surface radiation $S_{tot,~surf}
(S_W, f^\prime)$ according to (\ref{A2.5}), and the surface temperature,
given by (\ref{A2.10}), as well as the temperature difference consistently
also obtained from (\ref{C.14}).}
{\begin{tabular}{@{}cccc@{}} \Hline 
\\[-1.8ex] 
$S_W [Wm^{-2}] (f^\prime)$ & $c \left( \frac{S_W}{S_0}, f^\prime \right)$
& $S_{tot, surf} [Wm^{-2}]$ & $T_{surf} [K]$ \\
\Hline
43 & 1.67159 & 401.182 & 290.028 \\
$(f^\prime = 0.100)$ & & & \\
43 & 1.68250 & 403.800 & 290.500 \\
$(f^\prime = 0.092015)$ & & & \\
&  & $\Delta S_{up} = \Delta_{2 \times CO_2} = 2.6$ & $\Delta T = 0.47$ \\
\Hline
\end{tabular}}
\label{tabB2}
\end{table}

The numerical results in Table B2 are based on the input parameters, $S_0
= 240~Wm^{-2}$ and $S_W = 43 Wm^{-2}$, as well as $f^\prime = 0.1$ and
$f^\prime = 0.1 + \Delta f^\prime = 0.092015$, as obtained by evaluation
of (\ref{C.13}) for $\Delta_{2 \times CO_2} = 2.6 Wm^{-2}$ taken from
Table 6.

In a first step, upon evaluating (\ref{A2.5}) and (\ref{A2.10}), we obtained
the surface radiation and surface temperature shown in the second and
third lines of Table B2. The subtraction of the two surface temperatures from
each other then yields $\Delta T = 0.47 K$ shown in the last line of
Table B2. In a second step, we confirm $\Delta T = 0.47 K$~\footnote{The
discrepancy between $\Delta T = 0.47 K$ in Table B2 and $\Delta T =
0.46 K$ in Table 8 of the main text is due to the different input of
$T = 293 K$ in the main text.} by evaluating (\ref{C.14}) by either
inserting $\Delta f^\prime$ from (\ref{C.13}) or by directly inserting
$\Delta_{2 \times CO_2} = 2.6 Wm^{-2}$.


\section*{References}

\begin{thebibliography}{0}
\bibitem{Schack_PB}
A. Schack, 
{\it Phys. Bl{\"a}tter} {\bf 28} 26 (1972).

\bibitem{Schack}
A. Schack, {\it Der industrielle W{\"a}rme{\"u}bergang} (Verlag Stahleisen m.b.H.
D{\"u}sseldorf, 1. Auflage 1929, 8. Auflage 1983).

\bibitem{Gervais}
F. Gervais, 
Internat. Journal of Mod. Phys. B {\bf 28}, 1450095 (2014);\hfill\break
F. Gervais, {\it L'urgence climatique est un leurre} (Editions de
l'Artilleur, 75008 Paris, 2018).

\bibitem{Trenberth}
K.E. Trenberth, J.T. Fasullo and J. Kiehl,
{\it Bull. Am. Meteorol. Soc.} {\bf 90} 311 (2009).

\bibitem{HITRAN}
HITRAN data base.

\bibitem{Wei}
Peng-Shen Wei et al., {\it Heliyon} {\bf 4}, e00785 (2018).

\bibitem{Clark}
R. Clark, Ventura Photonics,
private communication, see Appendix A.

\bibitem{Clark2}
R. Clark,
{\it Energy and Environment} {\bf 24}, 319 and 341 (2013),
Ventura Photonics Monograph VPM 005 (2019).

\bibitem{Chilinger}
G.V. Chilinger et al.,
{\it Atmospheric and Climate Sciences} {\bf 4}, 819 (2014);
\hfill\break
O.G. Sorokhatin et al., {\it Global Warming and Global Cooling: Evolution of
Climate on Earth} (Elsevier 2007).

\bibitem{Harde}
H. Harde, 
{\it Internat. Journal of Atmospheric Sciences} {\bf 2017}, Article ID
9251034.

\bibitem{IPCC}
{\it Fifth Assessment Report of the Intergovernmental Panel on Climate Change
(IPCC), AR5}, ed. T.F. Stocker et al., Climate Change 2013; The Physical
Science Basis, (Cambridge University Press, New York, NY, USA, 2014).

\bibitem{Gervais_ESR}
F. Gervais, 
{\it Earth Sciences Reviews} {\bf 155}, 129 (2016).

\bibitem{Lindzen_Chou}
R.S. Lindzen, M.-D. Chou and A.Y. Hou,
{\it Bull. Am. Meteorol. Soc.} {\bf 82}, 417 (2001).

\bibitem{Lindzen_Choi}
R.S. Lindzen and Y.-S. Choi,
{\it Geophys. Res. Lett.} {\bf 36}, L16705 (2009).

\bibitem{Barkstrom}
B.R. Barkstrom,
{\it Bull. Am. Meteorol. Soc.} {\bf 65}, 1170 (1984).

\bibitem{Wielicki}
B.A. Wielicki et al.,
{\it IEEE Trans. Geosci. Remote Sens.} {\bf 36}, 1127-1141 (1998).

\bibitem{Chung}
E.-S. Chung, B.J. Soden, B.-J. Sohn,
{\it Geophys. Res. Lett.} {\bf 37}, L10703 (2010).

\bibitem{Murphy}
D.M. Murphy et al.,
{\it Geophys. Res. Lett.} {\bf 37}, L09704 (2010).

\bibitem{Trenberth_Fasullo}
K.E. Trenberth, J.T. Fasullo, C. O'Dell and T. Wong,
{\it Geophys. Res. Lett.} {\bf 37}, L03702 (2010).

\bibitem{Lindzen_Choi_new}
R.S. Lindzen and Y.-S. Choi,
{\it Asia-Pacific J. Atmos. Sci.} {\bf 47 (4)}, 377 (2011).

\bibitem{Bates}
J.R. Bates,
{\it Earth and Space Science} {\bf 3}, 207 (2016).

\bibitem{Kauppinen}
J. Kauppinen and P. Malmi, ``No experimental evidence for the significant
anthropogenic climate change'',
arXiv: 1907.00165v1 [physics.ao-ph].

\bibitem{Tsonis}
A. A. Tsonis, {\it Atmospheric Thermodynamics}, (2nd Edition,
Cambridge University Press, Cambridge UK, 2007).



\end{thebibliography}
\end{document}